\documentclass[apj]{emulateapj}

\usepackage{graphicx}

\usepackage[backref,breaklinks,colorlinks,citecolor=blue]{hyperref}
\usepackage{amsmath}
\usepackage{amssymb}

\shorttitle{Transient IR emission from tidal disruption flares}
\shortauthors{van Velzen et al.}

\begin{document}

\title{Discovery of transient infrared emission from dust \\ heated by stellar tidal disruption flares}

\author{S. van Velzen\altaffilmark{1}, A.\,J. Mendez, J.\,H. Krolik}

\affil{Department of Physics \& Astronomy, The Johns Hopkins University, Baltimore, MD 21218, USA}

\and
\author{V. Gorjian}
\affil{Jet Propulsion Laboratory, California Institute of Technology, Pasadena, CA 91109, USA}

\altaffiltext{1}{Hubble Fellow, \texttt{sjoert@jhu.edu}}

\begin{abstract}
Stars that pass within the Roche radius of a supermassive black hole will be tidally disrupted, yielding a sudden injection of gas close to the black hole horizon which produces an electromagnetic flare. A few dozen of these flares have been discovered in recent years, but current observations provide poor constraints on the bolometric luminosity and total accreted mass of these events. 
Using images from the Wide-field Infrared Survey Explorer (WISE), we have discovered transient 3.4~$\mu$m emission from several previously known tidal disruption flares. The observations can be explained by dust heated to its sublimation temperature due to the intense radiation of the tidal flare. From the break in the infrared light curve we infer that this hot dust is located $\sim 0.1$\,pc from the supermassive black hole. Since the dust has been heated by absorbing UV and (potentially) soft X-ray photons of the flare, the reprocessing light curve yields an estimate of the bolometric flare luminosity. For the flare PTF-09ge, we infer that the most likely value of the luminosity integrated over frequencies at which dust can absorb photons is ${8\times 10^{44}}$\,erg~s$^{-1}$, with a factor of 3 uncertainty due to the unknown temperature of the dust. This bolometric luminosity is a factor $\sim 10$ larger than the observed black body luminosity. Our work is the first to probe dust in the nuclei of non-active galaxies on sub-parsec scales. The observed infrared luminosity implies a covering factor $\sim1$\% for the nuclear dust in the host galaxies. 
\end{abstract}

\maketitle 

\section{Introduction}

Studying the inner parsec of galaxies with imaging observations is notoriously difficult due to the very high concentration of stars and the finite resolution of our telescopes. Fortunately, reverberation of time-variable signals can provide information on scales much smaller than can be spatially resolved. The response of broad emission lines to variability in the continuum light of active galactic nuclei (AGN), for example, yields a scale for their emission region of light days to light years \citep{Kaspi00}. Hot dust in a torus around the AGN disk can reprocess the optical/UV flux from the disk, re-emitting it at a typical temperature $T\approx 1500$~K \citep{Barvainis87}. Indeed the infrared (IR) light curve is observed to lag the optical light curve and implies a torus size of $\sim 0.1\, (L_{X}/10^{44} {\rm erg}\,{\rm s}^{-1})^{1/2}$~pc  (with $L_{X}$ the hard X-ray luminosity of the AGN, \citealt{Koshida14}).

Similar to reverberation mapping of AGN, gas at the centers of inactive galaxies can be studied via light echoes produced after the central supermassive black hole tidally disrupts a star. A few dozen tidal disruption flares (TDFs) candidates have been discovered in the last decade, first using X-ray  \citep[e.g.,][]{KomossaBade99,Esquej08,Saxton12} and UV satellites \citep{Gezari06,Gezari09}, and more recently using optical surveys \citep[e.g.,][]{vanVelzen10,Gezari12,Chornock14,Arcavi14,Holoien14,Holoien15}.

Light echoes from TDFs could be detected by monitoring (broad) emission lines in optical spectra of these flares \citep{Komossa08}. However the origin of observed TDF emission lines is unclear: the lines may not be due to photoionization of the pristine gas around the black hole, but may instead originate from within the photosphere of the stellar debris \citep{Roth15}. 

When a stellar tidal disruption results in the launch of a powerful relativistic jet \citep{Zauderer11,Bloom11,Burrows11,Levan11,Cenko11}, the radio light curve of this jet can be used to estimate the density of the circumnuclear gas as a function of distance to the black hole \citep{Berger12}. Yet this inference requires the assumption that the microphysical parameters of the jet (e.g., the fraction of energy in magnetic fields) remain unchanged as it expands. 

A new, and hitherto untested, use of TDFs to study the nuclei of inactive galaxies is transient IR emission from dust heated by the intense radiation field of the flare \citep{Lu16}.  

The typical scales relevant to IR reprocessing can be estimated from existing data and dust grain physics. The few dozen TDF {candidates discovered so far are bright ($L \sim 10^{43.5}~{\rm erg}\,{\rm s}^{-1} $) month-long flares that can be described by a single-temperature black body at optical/UV frequencies, $T_{\rm BB} \sim {\rm few } \times 10^{4}$~K \citep{vanVelzen10,Gezari12},  or soft X-ray frequencies, $kT \sim 0.05$~keV \citep{KomossaBade99,Miller_14li}. This high luminosity implies any dust near the black hole will quickly evaporate. The dust temperature is set by the flux absorbed by the grains and thus decreases with increasing distance to the black hole. For graphite grains with a radius of 0.1~$\mu$m, we find 
\begin{equation}
R \approx 0.15 (L_{\rm abs}/10^{45} {\rm erg\,{\rm s}^{-1}})^{1/2} (T_{\rm d}/1850~K)^{-2.9}~{\rm pc} 
\end{equation}
\citep{Barvainis87,Waxman00}.
Here $T_{\rm d}$ is the dust temperature and $L_{\rm abs}$ is the flare luminosity integrated over the frequencies where dust absorbs.  We normalized $T_{\rm d}$ to the expected sublimation temperature of the dust (discussed in detail in Sec.~\ref{sec:theo}). If 1\% of the radiated TDF energy ($E_{\rm TDF}\sim 10^{51}$~erg) is reprocessed and re-emitted over one year, the expected IR luminosity is $3\times 10^{41}\,{\rm erg}\,{\rm s}^{-1}$ or $\sim 1$\% of the typical galaxy luminosity in the $K$-band. 

To search for a dust reprocessing signal from TDFs, we used Wide-field Infrared Survey Explorer \citep[WISE;][]{Wright10} images at 3.4 and 4.6~$\mu$m. Besides a very stable photometric performance, an advantage of using WISE observations for this search is a factor of 2 reduced host galaxy flux at 3.4~$\mu$m compared to the $K$-band. 

Our work is the first\footnote{ Two preprints on TDF dust reprocessing \citep{Dou16,Jiang16} appeared a few days after the preprint of this manuscript was posted to the arXiv. Our discovery of dust reprocessing in WISE data was made public in \cite{vanVelzen15_Spitzer}.} to present transient IR emission from thermal (i.e., non-relativistic) TDFs, showing that dust heated by a TDF is detectable. These observations thus open a new wavelength regime for the study of TDFs and provide a new probe to study dust in galaxy centers. 

Prior to the work presented here, IR observations of TDFs have been focused mainly on the relativistic TDF Swift~J1644+57 \citep{Bloom11,Zauderer11,Burrows11,Levan11}, whose bolometric output is dominated by synchrotron emission from a relativistic jet. The $K$-band and {\it Spitzer} 3.6 and 4.5~$\mu$m emission of Swift J1644+57, observed about 300 days after discovery \citep{Levan16,Yoon15}, appear to fall on the Rayleigh-Jeans tail of a blackbody with a temperature $\gtrsim 10^{4}$~K \citep{Lu16}. This thermal component could be due to a stellar explosion or an accretion disk \citep{Levan16}, but the late-time IR emission of Swift~J1644+57 is not consistent with dust reprocessing.

The TDF candidate SDSS~J0952+2143, identified in the Sloan Digital Sky Survey \citep[SDSS;][]{york02} spectroscopic sample based on its extreme coronal emission lines \citep{Komossa08}, has been followed-up with {\it Spitzer} mid-IR spectroscopic observations to search for signs of dust heated by the flare. These observations were obtained about 4 years after the peak of the optical light curve \citep{Palaversa16}. WISE observations obtained 6 years after the peak suggest this {\it Spitzer} flux has faded by a factor of two \citep{Dou16}, providing strong evidence for transient IR emission due to dust reprocessing.

The outline of this paper is as follows. We first present the necessary theoretical background for dust reprocessing in the context of TDFs in Section~\ref{sec:theo}. We then discuss the steps of our data analysis in Section~\ref{sec:ana}: sample selection (Sec.~\ref{sec:parent}), co-addition of WISE images (Sec.~\ref{sec:co-add}), and photometry (Sec.~\ref{sec:photo}). In Section~\ref{sec:fit}, we explain how the parameters of our dust reprocessing model are estimated from the IR light curves. We discuss the results in Section~\ref{sec:results} and close with a list of our conclusions. We adopt a flat $\Lambda$CDM cosmology with $\Omega_{\Lambda}=0.7$ and $H_{0}=70$~km\,s$^{-1}$

%
\section{Dust reprocessing for tidal disruption flares}\label{sec:theo}
%
Because hydrogen ionization can remove photons before they heat the dust, we first estimate the magnitude of this effect and find that both H photoionization and H$_{2}$ photodissociation absorb only a very small fraction of the TDF light. 

In the initial stages of the TDF, its optical/UV continuum rises steeply toward a maximum.   As these first photons travel outward through surrounding gas, they ionize its H atoms. The spectra of most TDF are reasonably
well-fit by blackbodies of constant temperature $\simeq 3 \times 10^4$~K \citep{Gezari09,vanVelzen10,Arcavi14,Chornock14,Holoien15}. In such a case, the ratio of the number of ionizing photons radiated over the course of the flare to the number of nearby H atoms is
\begin{equation}
\frac{N_{\rm ion}}{N_{\rm H}} \simeq 6\times 10^{3} E_{50} R_{\rm 0.1}^{-3} n_{\rm H,3}^{-1},
\end{equation}
where $E_{50}$ is the integrated optical/UV energy in units of $10^{50}$~erg, $R_{\rm 0.1}$ is the distance of the ionized region in units of 0.1~pc, and $n_{\rm H,3}$ is the volume density of H atoms in units of $10^3$~cm$^{-3}$.   Thus,  the flare is more than capable of ionizing all the atoms out to nearly a parsec unless the density is $\gtrsim 10^6$~cm$^{-3}$.  The availability of ionizing photons diminishes for lower blackbody temperatures, but drops by only a factor of 30 if the temperature is as low as $1.5 \times 10^4$~K. If the gas is initially molecular rather than atomic, the relevant photons are in the 7.5--13.6~eV range, which are even more abundant in the light radiated by the flare than those above the ionization edge, so H$_{2}$ molecules will be broken into separate atoms even more quickly than individual H atoms are ionized.  Moreover, because the recombination time is $\simeq 1.6 \times 10^2 n_{e,3}^{-1}$~year, these atoms stay ionized for at least several years unless $n_e \gtrsim 10^5$~cm$^{-3}$. 

The flare is similarly able to evaporate the nearer dust grains. The binding energy per atom is 5.7~eV for astronomical
silicate or 6.8~eV for graphite \citep{Guhathakurta89}. If the total number density of C, N, O, and Si atoms is $\simeq
1.0 \times 10^{-3}n_{\rm H}$ \citep{Asplund09} and half of them are in grains, the ratio of radiated flare energy to grain
binding energy is
\begin{equation}
\frac{E_{\rm flare}}{E_{\rm binding}} \simeq 1.5 \times 10^8 E_{50} R_{\rm 0.1}^{-3} n_{\rm H,3}^{-1} \quad.
\end{equation}
This order of magnitude estimate shows that sublimating all dust grains within 0.1~pc would remove only a tiny fraction of the TDF photons.

While there is enough energy in the optical/UV flare to vaporize the grains out to parsec distances, the time required for grains to sublimate can be significant. Following \citet{Guhathakurta89}, we find the grain survival time, 
\begin{align}
t_{\rm sub} &= \frac{a}{da/dt} \nonumber \\ 
						&\simeq 3.21\, a_{0.1} \exp\left[42.747(1900/T_{\rm d} - 1)\right] \hbox{~month} \label{eq:Tsub}
\end{align}
where $T_{\rm d}$ is the grain temperature, the factor 42.747 applies to graphite grains, and we have normalized the grain size to 0.1~$\mu$m.  

As the dust evaporates, the IR flux fades rapidly. For a single dust grain, 
\begin{equation}\label{eq:Ldust}
L_{\rm dust} = 4\pi a^{2}\, Q_{\rm IR}  \sigma T_{\rm d}^{4} \quad ,
\end{equation}
with $Q_{\rm IR}$ the emission efficiency. At the wavelength of our observations
and for grains with $a \gtrsim 0.1$~$\mu$m,  $Q_{\rm IR}\propto a^{2}$ \citep[cf. Figure 4b in ][]{Draine84}. Since $L_{\rm dust} \propto a^4 $, we have
\begin{equation}
\frac{L} {dL/dt} = \frac{1}{4}\frac{a}{da/dt}
\end{equation}
and we find the effective sublimation temperature ($T_{\rm sub}$) by equating one fourth of the sublimation time for a given grain (Eq.~\ref{eq:Tsub}) to the duration of the TDF. The full width at half maximum (FWHM) of the optical TDF light curve of \mbox{PTF-09ge} is $\Delta t_{\rm opt} \approx3$ months, which yields $T_{\rm sub}=1850$~K. At this temperature, silicate dust sublimates in $\sim 1$~hr, hence this grain type can survive only at larger radii from the black hole compared to graphite. As we will show below, a large radius leads to a lower IR luminosity because the reprocessed energy is emitted over a longer time. We therefore model the reprocessing signal using only graphite grains. The sublimation temperature is weakly dependent on grain size, e.g., for $a =0.01$~$\mu$m, the sublimation temperature is 100~K lower.

Following estimates of the dust heated by AGN or GRBs \citep{Barvainis87,Waxman00}, we compute the radius of a dust shell as a function of its temperature by equating the rate at which the grains absorb heat, 
\begin{equation}
Q_{\rm UV} \frac{L_{\rm abs} a^{2}}{4R^{2}} \quad,
\end{equation}
to the rate at which dust radiates (Eq.~\ref{eq:Ldust}): 
\begin{equation}
R = \left(\frac{L_{\rm abs}}{16\pi \sigma_{\rm SB} T_{\rm d}^{4}} \frac{Q_{\rm UV}}{\left<Q_{\rm IR} \right>_{T}} \right)^{{1/2}}  \quad .
\end{equation}
Here $Q_{\rm UV}\approx1$ is the absorption efficiency of the grains and $\left< Q_{\rm IR} \right>_{T}$ is the temperature-averaged emission efficiency. The latter depends on the shape of the dust spectral energy distribution (SED). Using a modified black body spectrum, $B'(T)= \nu^{q}B(T)$, with $q=1.8$ at the wavelength of our observations \citep{Draine84}, the emission efficiency is
\begin{equation}
\left< Q_{\rm IR} \right>_{T} = 0.16 ~ a_{0.1}^{2} T_{1850}^{1.8}  \quad 
\end{equation}
\citep[][Eq.~6.1]{Draine84} and we thus find the radius of the dust shell as function of the dust temperature, size and luminosity of the heat source:
\begin{equation}\label{eq:R_sub}
R = 0.15 \left(\frac{L_{\rm 45}} { a_{0.1}^{2}T_{1850}^{5.8}}\right)^{1/2}~{\rm pc} \quad.
\end{equation}
Here $L_{\rm 45}=L_{\rm abs}/10^{45}\,{\rm erg}\,{\rm s}^{-1}$ and $T_{1850}=T_{\rm d}/1850$~K. For $T_{1850}=1$, Eq.~\ref{eq:R_sub} yields the sublimation radius. Outside the sublimation radius, the grain temperature stays in close equilibrium with the incident radiation for the grain cooling time is extremely small:
\begin{equation}
t_{\rm cool} \simeq 1 \times 10^{-4} a_{0.1}^{-1} T_{1850}^{-5}~\hbox{~s}.
\end{equation}

As will be discussed below in Section~\ref{sec:fit}, the shape of the IR light curve can be used to estimate the radius from the black hole where this emission originates. We therefore rewrite Eq.~\ref{eq:R_sub} to find the luminosity of the flare as a function of this radius:
\begin{equation}\label{eq:L_from_R}
L_{\rm abs} = 5 \times 10^{44} ~ R_{\rm 0.1}^2 a_{0.1}^2 T_{1850}^{5.8}~{\rm erg}\,{\rm s}^{-1} \quad.
\end{equation}
From this expression we see that the flare luminosity inferred from the reprocessing light curve is sensitive to both the dust temperature and the dust grain size. The dust temperature can be measured using multi-frequency IR follow-up observations of TDFs; an accurate measurement of this temperature coupled with the flare duration would place a lower bound on the characteristic grain size.

For a typical distribution of grain sizes (${\rm d}n/{\rm d}a\approx a^{-3.5} $; \citealt{Weingartner01}), the largest graphite grains determine the effective sublimation radius because they dominate the luminosity at 3~$\mu$m: 
\begin{align}
L_{\rm IR} \propto&  \int_{\rm a_{\rm min}}^{\rm a_{\rm max}} {\rm d}a~L_{\rm dust} \frac{dn}{{\rm d}a} \propto  \int_{\rm a_{\rm min}}^{\rm a_{\rm max}} {\rm d}a~a^{2} Q_{\rm IR} a^{-3.5}  \nonumber  \\
\propto& \int_{\rm a_{\rm min}}^{\rm a_{\rm max}} {\rm d}a~a^{2} a^{2} a^{-3.5} \sim a_{\rm max}^{1.5}\quad.
\end{align}
The size normalization used in the expressions for $R$ and $L_{\rm abs}$ (i.e., $a=0.1$~$\mu$m), is motivated by dust models for the extinction curve of the Milky Way and the Magellanic Clouds \citep{Weingartner01}, which yield a power-law dust-size distribution with an exponential cutoff at $a_{0.1}\approx1$.

The grains heated by the flare, but not evaporated, reradiate the energy absorbed in the near-IR.  However, because
they are distant from the black hole, there is a significant delay before the IR reaches a distant observer.  To
be specific, if the direction from the black hole to the observer is the polar axis of a system of spherical coordinates, the
delay is given by
\begin{equation}
\tau = (R/c)(1 - \cos\theta),
\end{equation}
where $R$ is the radial coordinate of a particular dust grain and $\theta$ is its polar angle. In response to an isotropically-radiated optical/UV flare with light curve $L(t)$, surrounding material produces an IR light curve
\begin{equation}
L_{\rm IR}(t) = \int \, d\tau \, \Psi(\tau) L(t- \tau),
\end{equation}
where $\Psi(\tau)$ is the response function.   For a spherical shell of material at radius $R_0$ responding linearly to the irradiating flux, 
\begin{align}
\Psi(\tau) & = \int \, d\phi \, \int \, d(\cos\theta) \, \int \, dR \, R^2 \frac{\partial j_{\rm IR}}{\partial L} \delta(R-R_0) \times \nonumber \\ & \phantom{----------}  \delta\left[\tau - (R/c)(1-\cos\theta)\right]  \nonumber \\
   & = 2\pi \frac{\partial j_{\rm IR}}{\partial L} \int_{-1}^{1} \, d (\cos\theta) \, R_0 c \delta\left[\cos \theta - (1 - c\tau/R_0)\right] \nonumber \\
   & = 2\pi R_0 c\frac{\partial j_{IR}}{\partial L}\label{eq:repdelta}
\end{align}
where $\frac{\partial j_{\rm IR}}{\partial L}$ is the marginal IR emissivity (when the dust radiates in the IR exactly the same energy it absorbs from optical/UV/X-ray flare light, the marginal IR emissivity is identical to the covering factor). The final value is obtained {\it only} for $\tau \leq 2 R_0/c$, otherwise the root of the $\delta$ function's argument lies outside the range of $\cos\theta$.   In other words, spherical shells of linear reprocessors generically produce square-wave response functions extending from $\tau=0$ to $\tau = 2R_0/c$.  

Our fiducial model is a thin and spherically symmetric shell of dust that is optically thin to its own IR emission. If $R_{\rm sub}\gg c \Delta t_{\rm opt}$, it predicts a square wave, independent of the detailed shape of the optical/UV light curve. 
In Section~\ref{sec:fit} we show that applying this transfer function to the TDF light curve provides a good description of the IR observations.

%
%
\section{Analysis}\label{sec:ana}
%
%
In the previous section we found that IR emission from dust heated by a TDF yields a signal that can be detected up to $\sim 1$ year after the peak of the flare. Below we present the details of our search for this signal using WISE images at 3.4 and 4.6~$\mu$m. 

We typically have two or three observations separated by 6 months in the period 2010--2012, and, thanks to the reactivation of WISE \citep{Mainzer14}, four more images from 2014 through 2015. Our parent sample (discussed in Sec.~\ref{sec:parent}) consists of optical TDFs. The flux of these targets was extracted from co-adds of the WISE exposures (discussed in Sec.~\ref{sec:co-add}) using a forced photometry method (discussed in Sec.~\ref{sec:photo}). The uncertainty on this flux was computed by repeating our photometry on a set of reference sources with a flux similar to the target, and thus includes both statistical and systematic sources of uncertainty. 

\begin{deluxetable}{l c c c c c c}
\tablewidth{0pt}
\tablecolumns{7}
\tablecaption{Parent sample of TDFs. \label{tab:parent}}
\tablehead{name & redshift & $t_{\rm peak}$ & $T_{\rm bb}$ & $\log L_{\rm bb}$ & $M_K$ \\
& & (year)  & ($\times 10^4~{\rm K}$) & (${\rm erg}\,{\rm s}^{-1}$) & (Vega)} \\
\startdata
PTF-09ge        & 0.0640 & 2009.4 & 2.2 & 44.1 & 14.8\\
PTF-09axc       & 0.1146 & 2009.6  & 1.2 & 43.5 & 15.5\\
D23H-1          & 0.1855 & 2007.7    & 4.9 & 44.0 & 16.0\\
TDE2            & 0.2560 & 2007.7    & 1.8 & 44.0 & 16.2\\
PTF-09djl       & 0.1840 & 2009.6    & 2.6 & 44.4 & 16.7
\enddata
\tablecomments{The third column,  $t_{\rm peak}$, lists the time of maximum light of the optical/UV TDF light curve (except for TDE2, which was observed post-peak and we use the first observation of this source). $T_{\rm bb}$ and $\log L_{\rm bb}$ denote the TDF black body temperature and luminosity, respectively. The last column, $M_{K}$, lists the total $K$-band magnitude of the host galaxy.}  
\end{deluxetable}

%
\subsection{Parent sample of TDFs}\label{sec:parent}
%
To be able to use the WISE observations of 2014--2015 as a baseline, we restricted our sample to TDFs that occurred between 2007 and 2010, leaving five optical/UV-selected flares (X-ray TDFs are excluded from the sample because their light curves are so sparsely sampled that the time of maximum light is poorly constrained). The properties of our five targets are listed in Table~\ref{tab:parent}, below we provide a brief review of these sources.

Three of the five TDFs in our parent sample were found in the Palomar Transient Factory \citep[PTF;][]{Law09} data by \citet{Arcavi14}: PTF-09ge, PTF-09djl, and PTF-09axc. Arcavi et al. selected relatively bright optical transients, $-21 < M_R < -19$. 
Optical spectra of the flares were obtained within a few months of the peak of the flare, covering the wavelength range 300-1000~nm (in the observer-frame). These spectra were used to estimate the black body temperature of these events. The optical spectra imply a post-starburst (or E+A) classification for the host galaxies of these three flares \citep{Arcavi14}. 

One source in our sample, TDE2, was found in SDSS  imaging observations by \citet{vanVelzen10}. This flare was also detected in GALEX imaging, about 1~year after the optical peak. The black body temperature was measured using SDSS ($u$, $g$, $r$, $i$) photometry. 

One source in our sample, D23H-1, was found by \citet{Gezari09} using the Galaxy Evolution Explorer \citep[GALEX;][]{Martin05} Time Domain Survey \citep{Gezari13}. This TDF was observed with {\it Chandra}, both 3 and 116 days after its peak; no source was detected in these X-ray follow-up observations, implying $L_{0.2-2~{\rm keV}}<10^{41}\,{\rm erg} \,{\rm s}^{-1}$. The black body temperature of this flare was computed from $g$-band, NUV and FUV photometry. 

The optical spectrum of the host galaxy of D23H-1 revealed narrow H$\alpha$ emission and a Balmer decrement. This implies ongoing star formation ($\approx 3\,M_{\odot}\,{\rm yr}^{-1}$) and a modest amount of extinction in the host galaxy: $E_{B-V} = 0.3$ \citep{Gezari09}. Correcting for this extinction, Gezari et al. found a factor 10 increase to the TDF black body luminosity relative to the luminosity derived from the uncorrected SED. It should be noted that the Balmer decrement is a galaxy-averaged property and provides a measurement of dust extinction toward the starforming regions where the Balmer lines are produced. The light from a TDF, on the other hand, samples dust in a narrow pencil beam to the galaxy center, hence the dust extinction derived from the Balmer decrement may not apply directly to TDFs. For this reason, and because the host galaxy spectra of the other TDFs in our sample do not allow an extinction measurement via the Balmer decrement, we will use the uncorrected black body luminosity of D23H-1. 

\begin{figure*}
\centering
\includegraphics[trim=7mm 10mm 0mm -5mm, width=0.19 \textwidth]{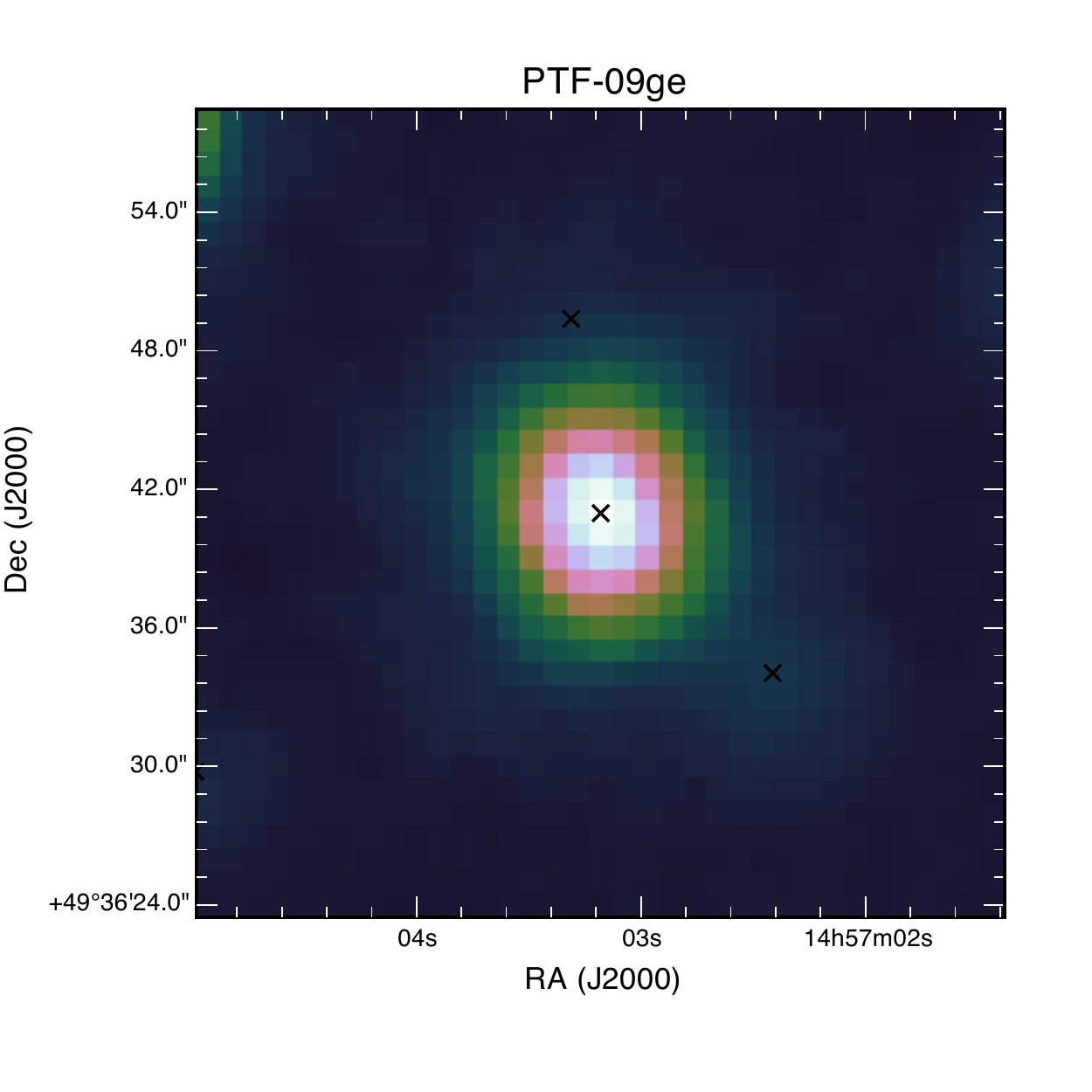}
\includegraphics[trim=7mm 10mm 0mm -5mm, clip, width=0.19 \textwidth]{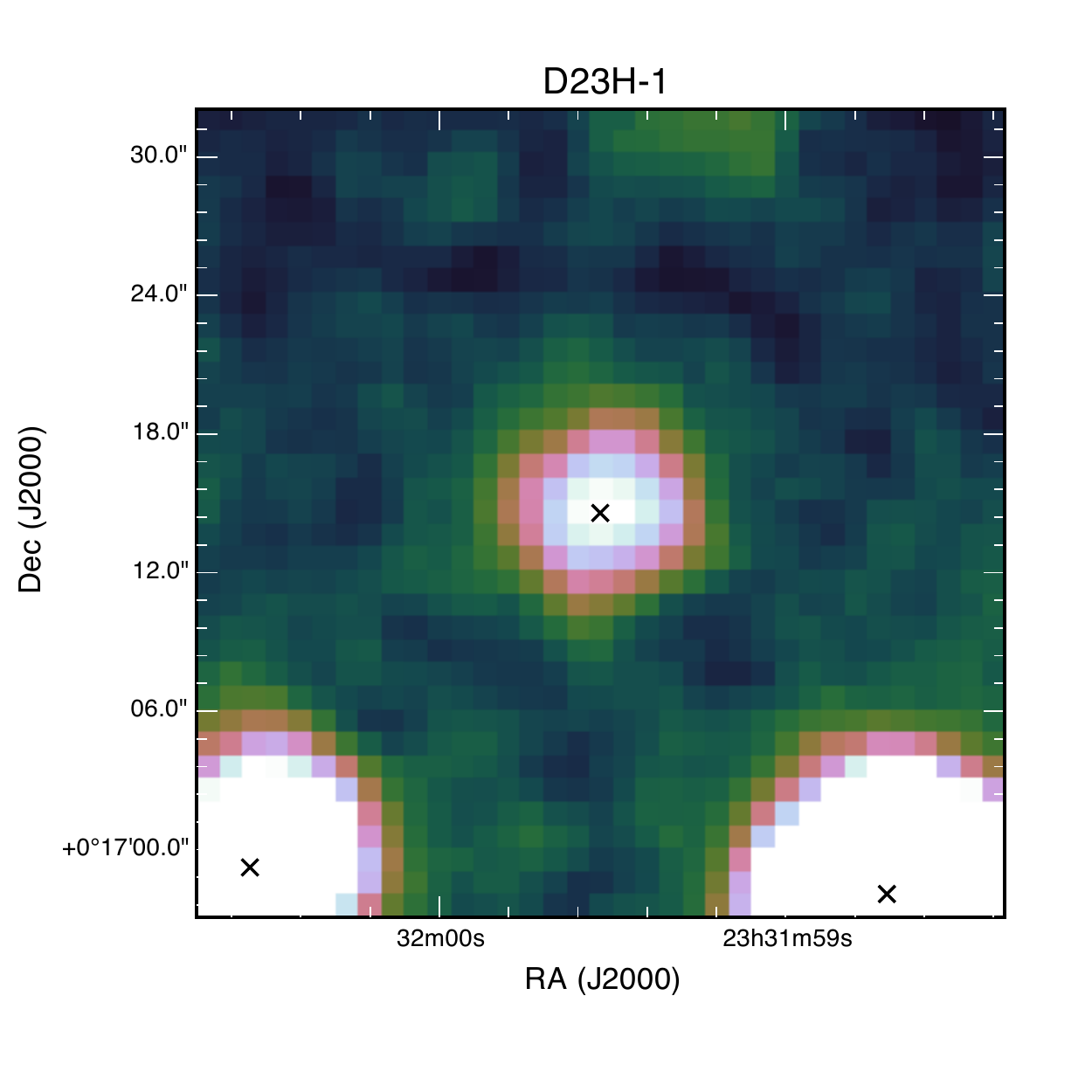}
\includegraphics[trim=7mm 10mm 0mm -5mm, width=0.18 \textwidth]{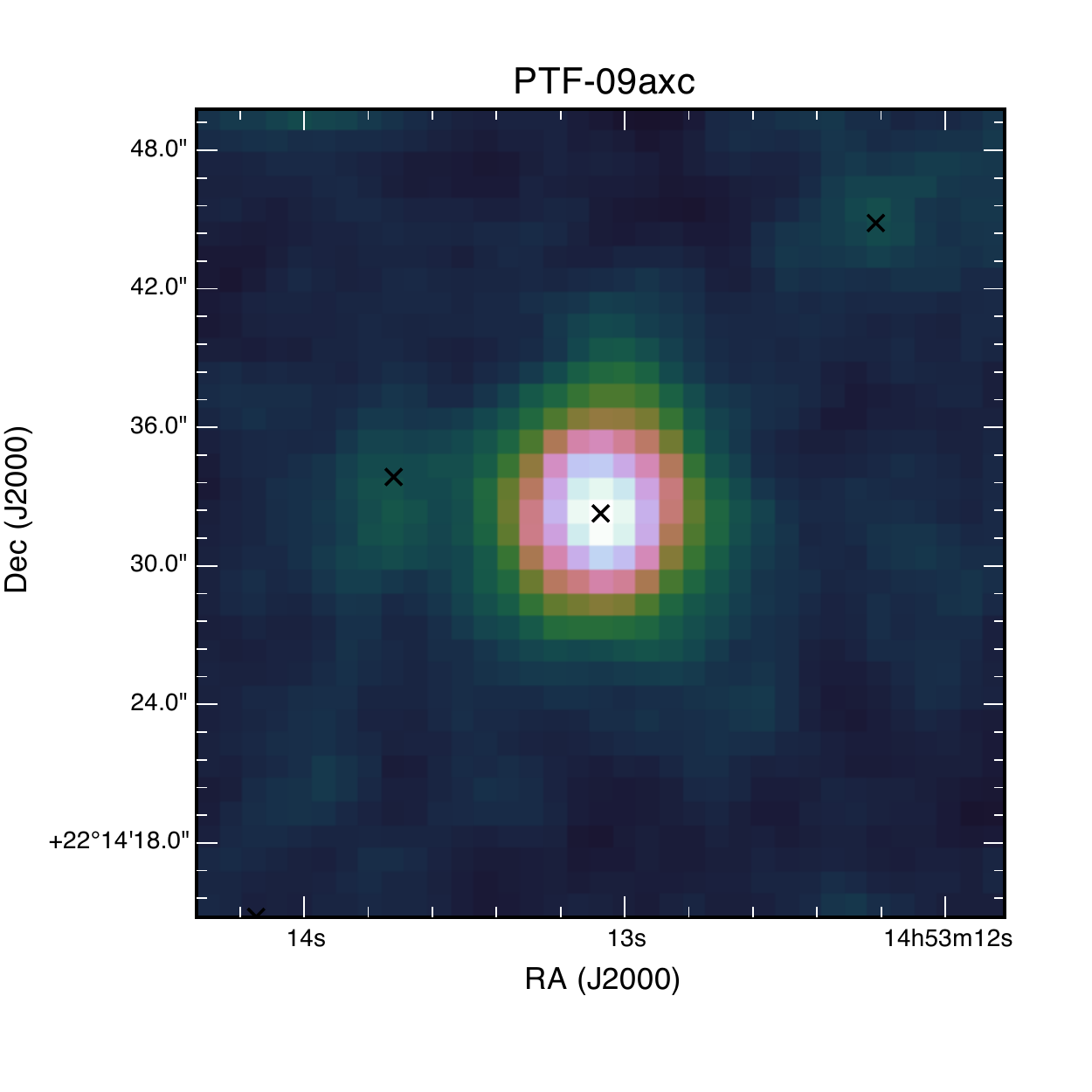} 
\includegraphics[trim=7mm 10mm 0mm -5mm, clip, width=0.19 \textwidth]{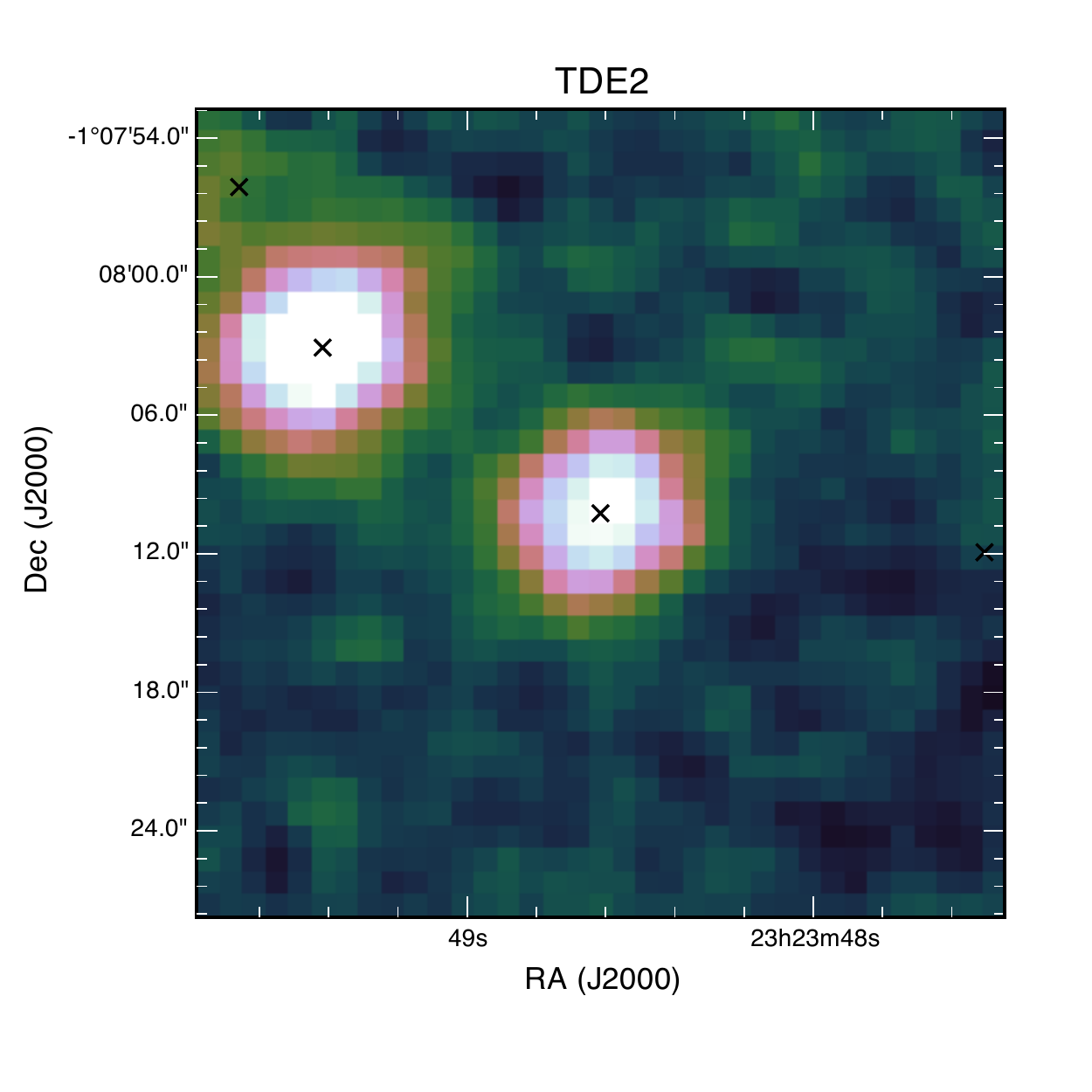}
\includegraphics[trim=7mm 10mm 0mm -5mm, clip, width=0.19 \textwidth]{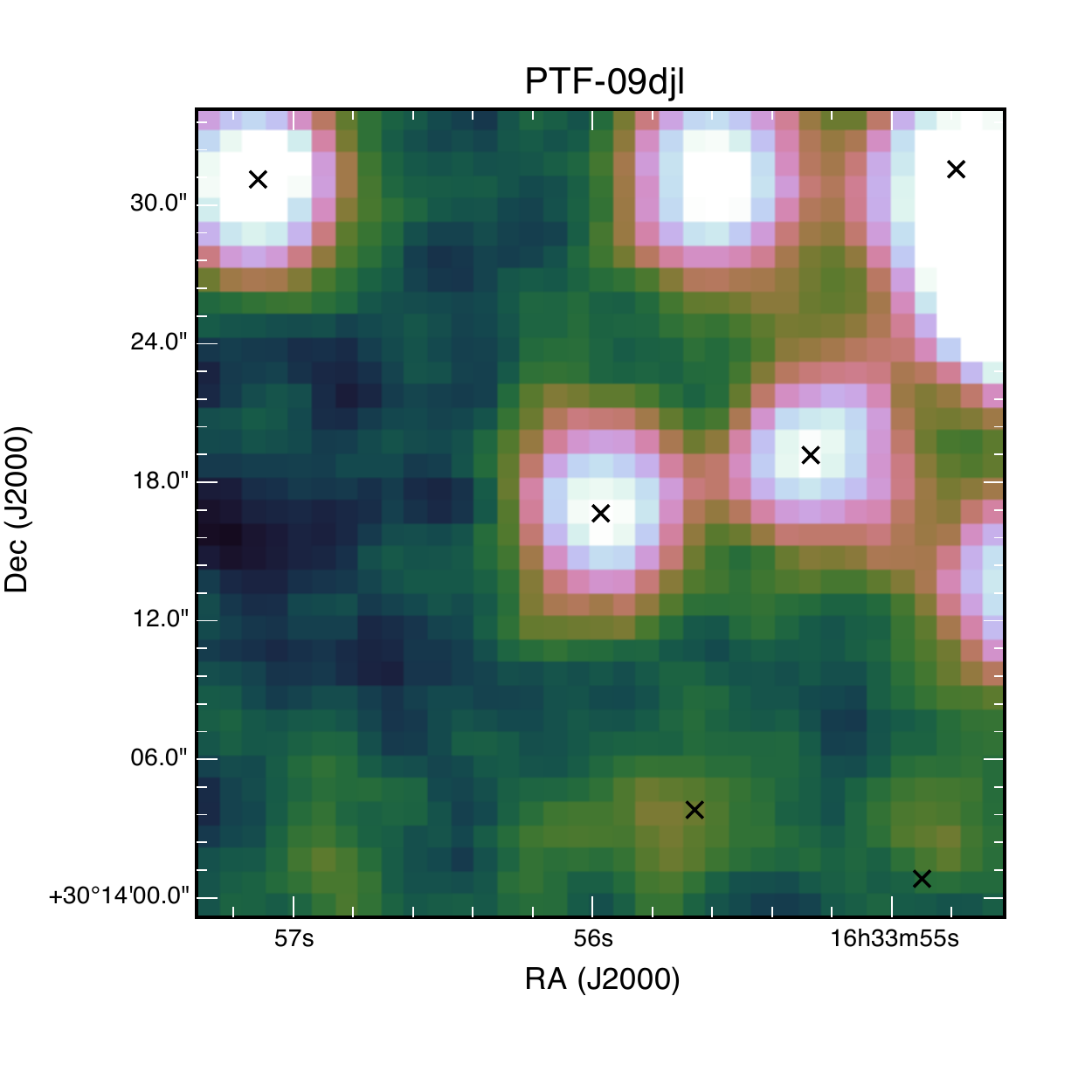} 
\caption{Cutouts of our WISE W1 co-adds,  $35"\times35"$ on a linear flux scale. Images are centered on the five TDFs in our parent sample. Our forced photometry method uses the coordinates of the unWISE catalog (black crosses) to measure the flux of the target simultaneously with other nearby sources.}\label{fig:cutout}
\end{figure*}

%
\subsection{Co-addition of WISE images}\label{sec:co-add}
%
The WISE satellite completed full sweeps of the sky in six months following great circles with the Sun at the center. There are about 15 orbits per day and most sources are observed 12 times in each scan \citep{Wright10}.  For the sources in our parent sample, we first investigated the WISE multi-epoch photometry catalog, which provides a profile flux measurement for each exposure in which a source is detected, finding significant variability for most of our targets. To obtain a higher signal-to-noise measurement, we co-added the individual exposures using the \textsc{icore} software package \citep{Masci13} --- which is similar to the software used by the WISE team to create their ``Atlas Images'' \citep{Masci09}.  

We produced co-adds of the individual ``level 1b'' images of the W1-band (3.4~$\mu$m) and W2-band (4.6~$\mu$m) --- the longer-wavelength bands (W3 and W4) are not available after September 2010, when the mission ran out of cryogenic coolant \citep{Mainzer11}. We excluded frames flagged for having low image quality. We adopted the default parameters of \textsc{icore} for WISE data and produced area-weighted co-adds with a pixel scale of 1.0" pix$^{-1}$ (compared to 2.75" pix$^{-1}$ for the input images). No ``drizzling'' was applied, i.e., the pixel size of the input images was kept fixed when mapped to the co-add frame. 

To have a sufficient number of reference sources in the field (see next section) we used a co-add size of $20' \times 20'$ (about a factor 2 smaller than the WISE field of view). The depth of coverage (i.e., number of individual images contributing to a co-add pixel) over this area is constant at the 5--10\% level; for a box of 10 pixels, the depth of coverage is constant at the 3--5\% level. 

For each co-add, we estimated the point-spread-function (PSF) by fitting a Gaussian ellipse to isolated stars \citep{Jones15}; we typically use 5--10 stars for this fit. The residuals of the fit are added to the Gaussian profile to find a good approximation of the true PSF. The typical PSF FWHM is 6". To enable relative photometry, we also co-added all individual exposures (i.e., all images from 2010 to 2015). Cutouts of these co-adds are shown in Fig.~\ref{fig:cutout}.

\begin{deluxetable}{lccccc}
\tablecaption{Summary of W1 photometry. \label{tab:photo}}
\tablewidth{0pt}
\tablecolumns{6}
\tablehead{name & \# of refs. & $\left<m\right>$ &$\left<\sigma(m)\right>$ & $\chi^{2}/{\rm dof}$ & $\log P(\chi^{2})$ \\ } \\ 
\startdata

PTF-09ge   &    34 & 16.96 & 0.016 & 62.7/7 & -10.9 \\
PTF-09axc  &    33 & 17.95 & 0.033 & 20.5/7 & -2.6 \\
D23H-1     &    32 & 18.51 & 0.044 & 18.9/6 & -2.7 \\
TDE2       &    41 & 18.44 & 0.073 &  8.1/6 & -0.8 \\
PTF-09djl  &    56 & 19.51 & 0.061 &  5.7/6 & -0.5 

\enddata
\tablecomments{For each TDF in our parent sample, we list the number of reference sources that were used by our relative photometry method (second column) and the mean W1 magnitude (third column). The mean of the rms scatter of the W1 light curve of the reference sources, $\left<\sigma(m)\right>$, measures the typical accuracy of our photometric method. The  $\chi^{2}$ over the degrees of freedom (dof) is computed for the W1 light curve of each TDF, using a model that has a constant flux as a function of time. The last column lists the probability to find the observed  $\chi^{2}/{\rm dof}$. }
\end{deluxetable}

%
\subsection{Photometry on WISE images}\label{sec:photo}
%
After obtaining a set of co-adds centered on each TDF, the next step is to extract the flux from these images. Given the depth and resolution of our co-adds, we anticipate blending of sources needs to be accounted for to obtain accurate photometry. Fortunately, all our targets are in the SDSS footprint, hence we can use information from the higher-resolution SDSS imaging to guide our WISE photometry. Our approach is motivated by the success of the unWISE project \citep{Lang14,Lang14b}, which used the measured SDSS source positions and star/galaxy profiles to construct a catalog of WISE photometry for 400 million SDSS sources.

To measure the flux of a given target, we place it at the center of a $35"\times 35"$ {\it scene}. We then measure the flux of all known SDSS sources in this scene by minimizing the difference between a model of these sources and the observed image using \textsc{galfit} \citep{Peng02}. For sources classified as stars by the SDSS pipeline \citep{stoughton02}, this model is simply given by the PSF of the co-add. For SDSS galaxies, we use a deVaucouleurs profile with parameters measured by the SDSS pipeline. Except for sources in the scene that are 2~mag brighter than the main target, we keep the centroid of the model fixed at the location measured by SDSS. By simultaneously fitting for the amplitude of all objects in the scene, we properly account for any contributions from nearby sources to the flux of the target. To obtain a stable solution, we fixed the amplitude of sources that are 2~mag fainter than the target at the magnitude reported by unWISE or sources with a distance from the center of the scene that is larger than 3 times the PSF FWHM. 

Our method of image co-addition leads to spatially correlated noise, hence the uncertainty obtained from fluctuations of the ``sky'' underestimates the true statistical uncertainty. To measure the uncertainty, we use a set of reference sources (both stars and galaxies) that are assumed to have a constant flux with time. We measure the flux of the reference sources using the \textsc{galfit} scene photometry described above. By comparing the observed reference source flux in each epoch, $m_{i}(t)$, to the flux in the co-add of all exposures, $\left< m_{i} \right>$, we can estimate the uncertainty on the flux in each epoch by fitting a Gaussian distribution to $m_{i} - \left< m_i \right>$. We also use the mean of this distribution to correct for any offsets in the zero point of the co-adds with respect to the zero point of the co-add of all exposures. The relative photometry light curves, as shown in Fig.~\ref{fig:alllc}, are thus given by
\begin{equation}\label{eq:relph}
m_{\rm rel, TDF}(t) = m_{\rm TDF}(t) - \left<m_{\rm TDF}\right> - \left< \,m_{i}(t) - \left< m_{i} \right>\, \right> \quad .
\end{equation}
With $i$ running over the reference sources. The relative zero point offsets, $\left< \,m_{i} - \left< m_{i} \right>\, \right>$, are found to be between 0.001~mag to 0.015~mag. To recover the absolute flux scale, we tied our measurement of the flux of the reference sources to the flux listed in the unWISE catalog.

\begin{figure}
\centering
\includegraphics[trim=8mm 12mm 3mm 1mm, width=0.39 \textwidth]{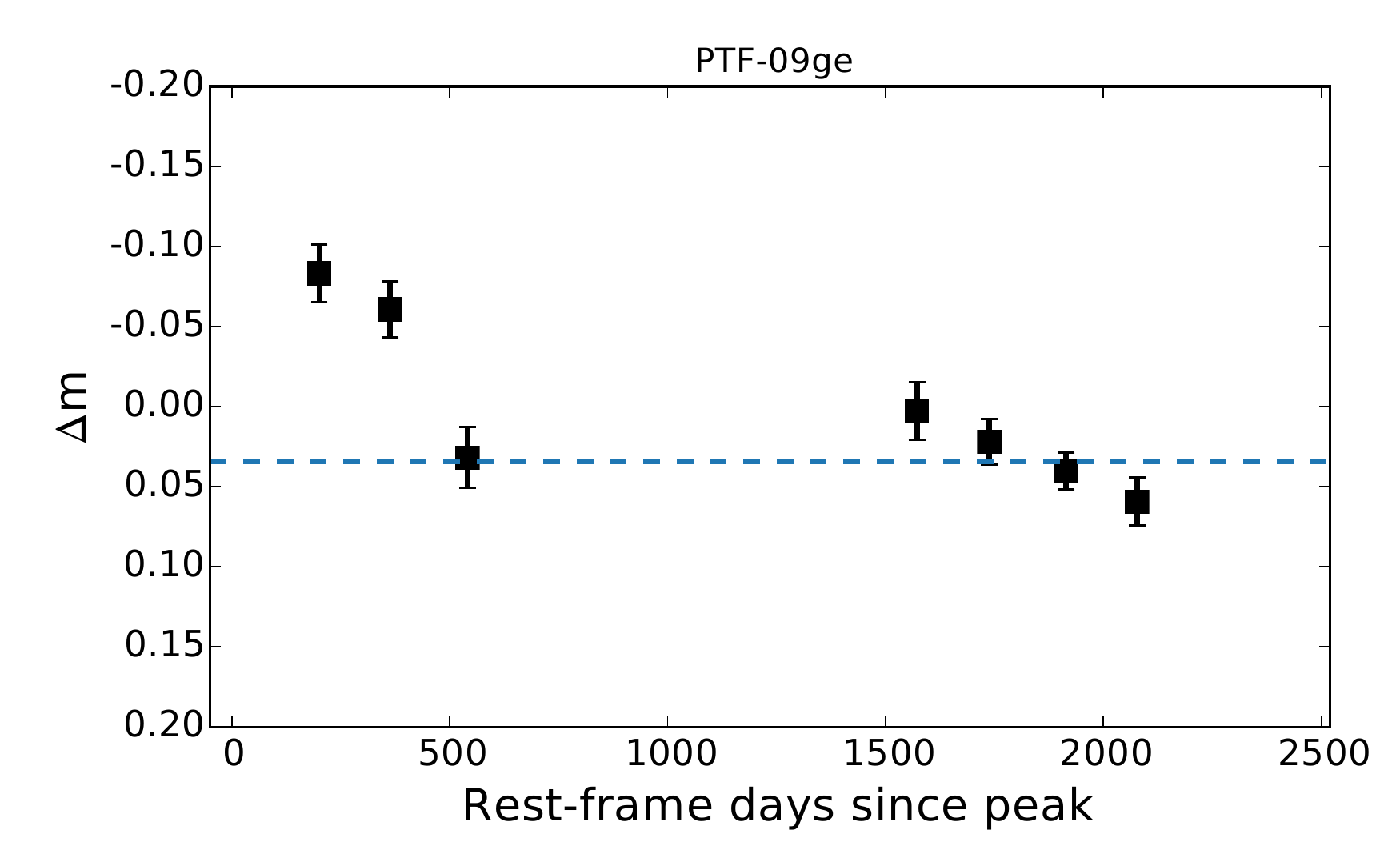}
\includegraphics[trim=8mm 12mm 3mm 1mm, width=0.39 \textwidth]{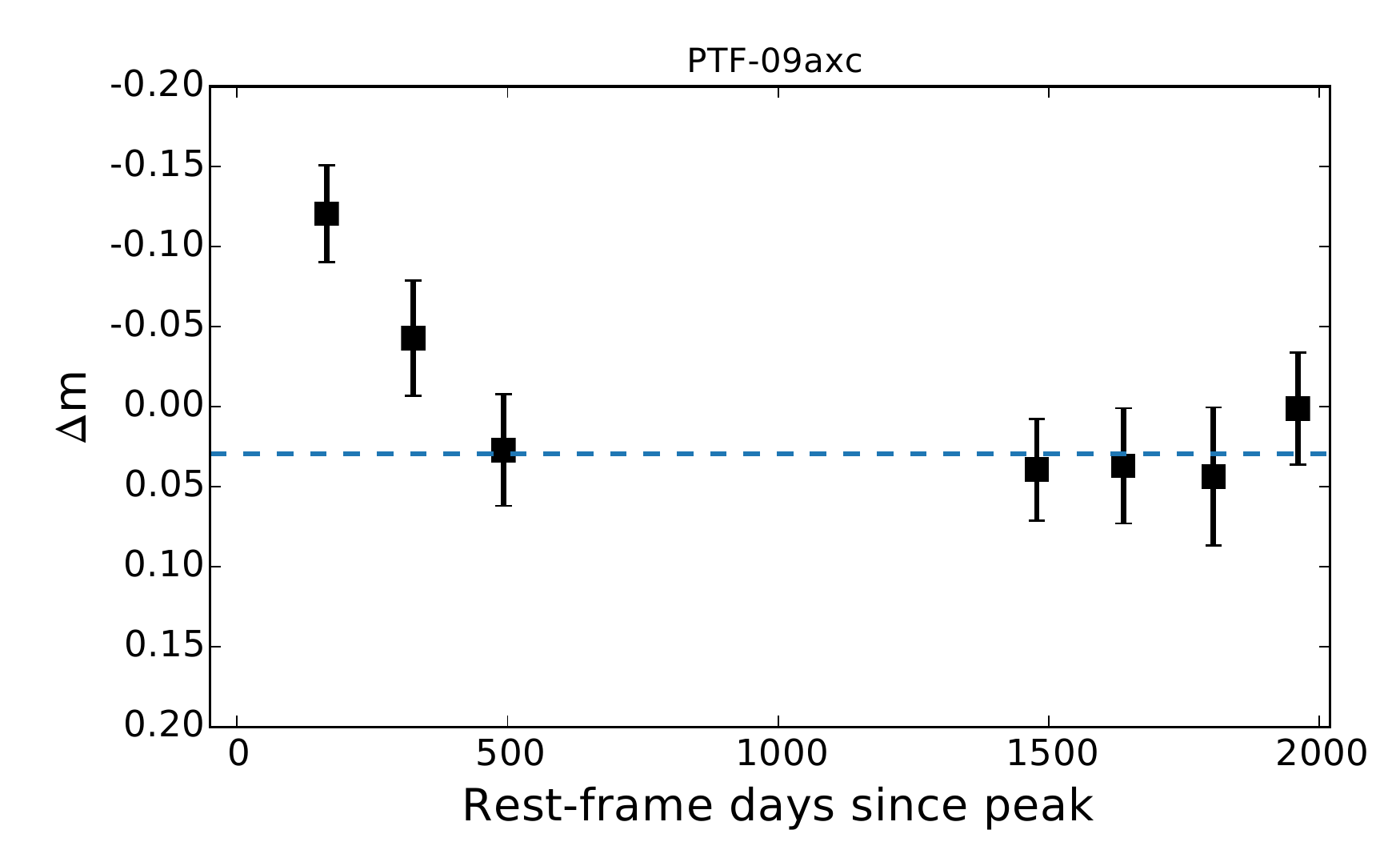}
\includegraphics[trim=8mm 12mm 3mm 1mm, width=0.39 \textwidth]{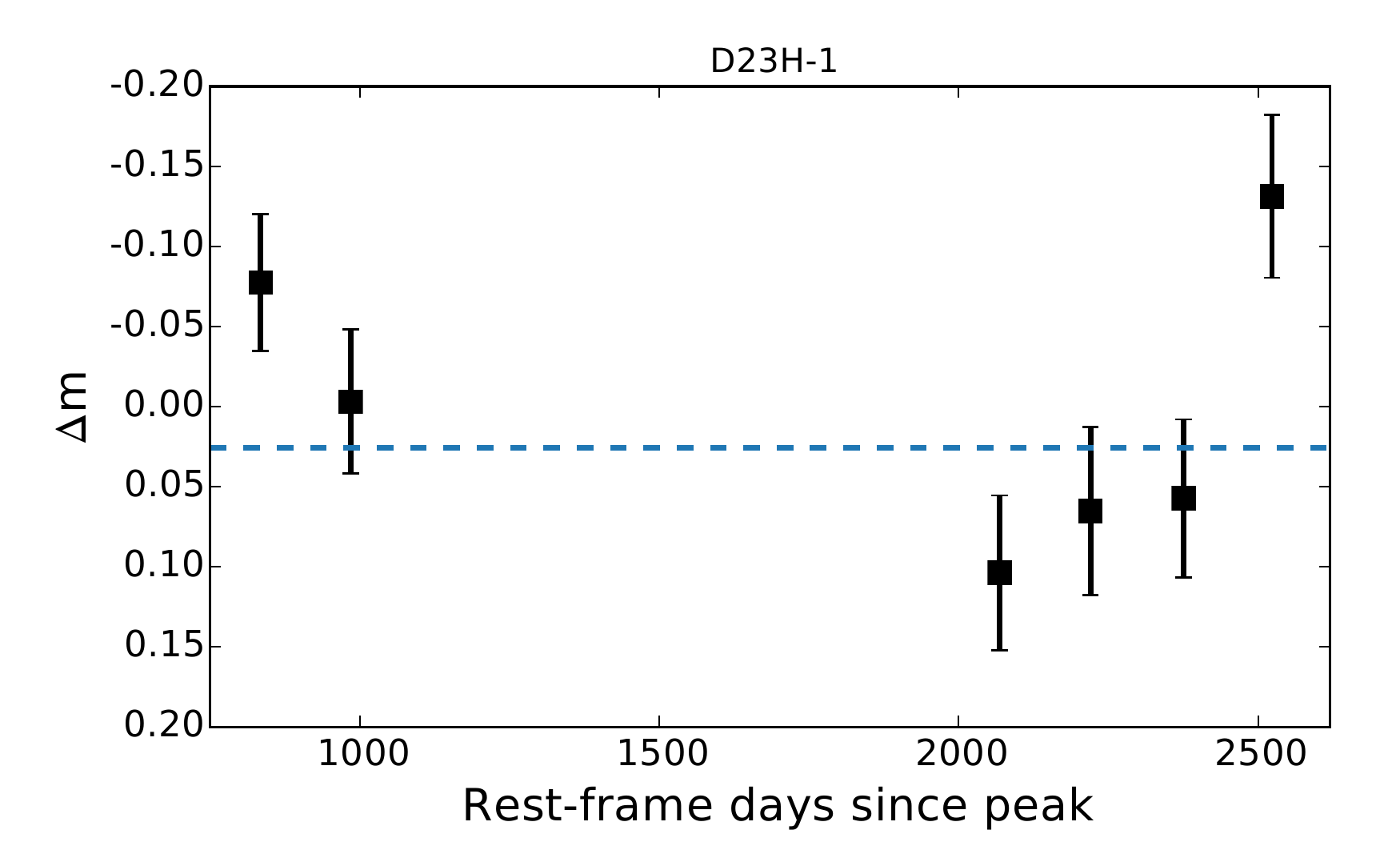}
\includegraphics[trim=8mm 12mm 3mm 1mm, width=0.39 \textwidth]{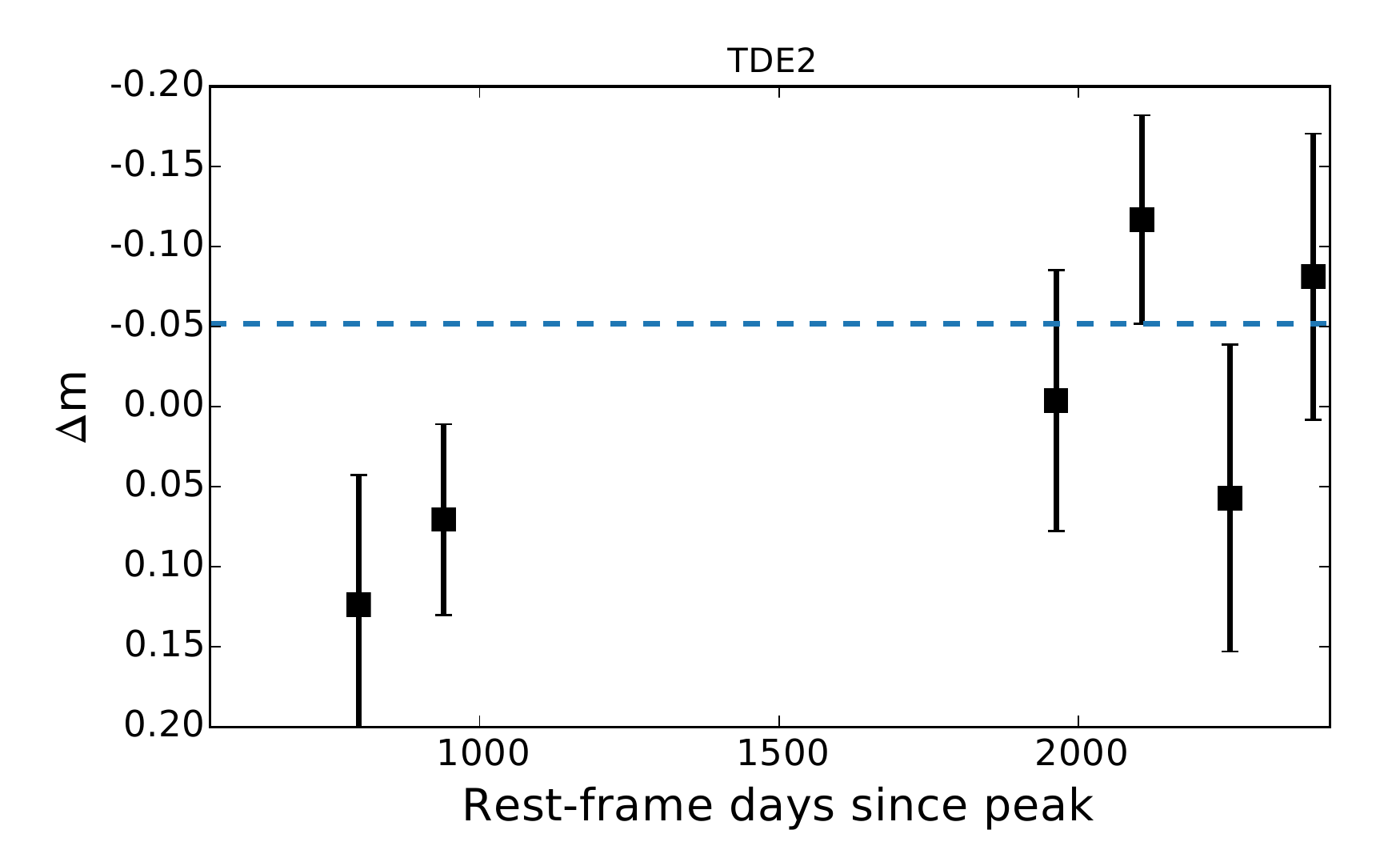}
\includegraphics[trim=8mm 12mm 3mm 1mm, width=0.39 \textwidth]{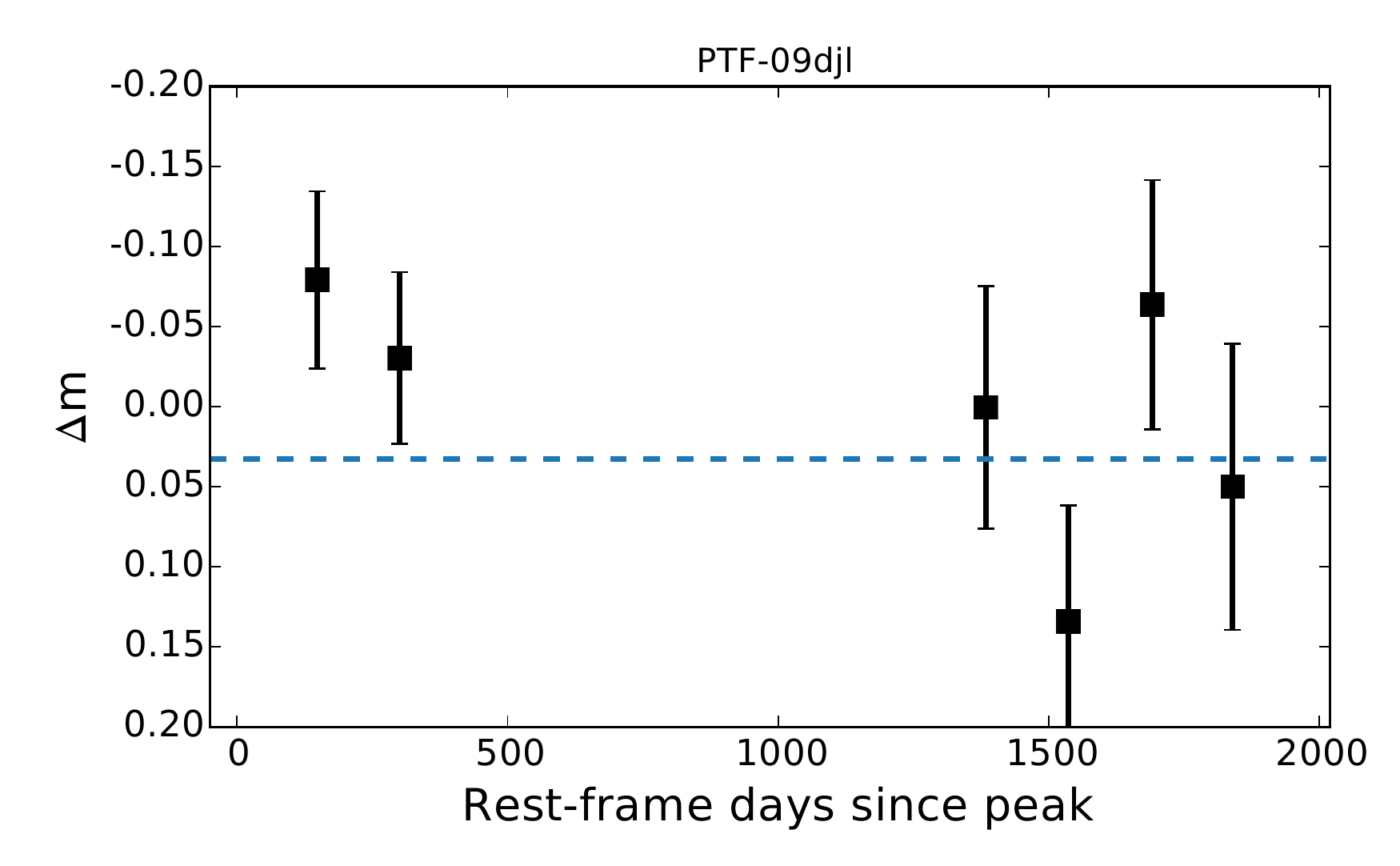} \\[10pt]
\caption{Relative photometry light curve at 3.4~$\mu$m of the five tidal disruption flares in our parent sample. The time is measured in days relative to the peak of the optical light curve. The uncertainty on the IR flux is computed from the root-mean-square variability of a set of reference sources with a similar IR flux as the target and thus accounts for both Poisson noise and systematics. The baseline level, estimated from the observations obtained in 2014 amd 2015, is indicated by the dashed line. The first three TDFs (PTF-09ge, PTF-09axc, D23H-1) show significant variability,  see Table~2.}\label{fig:alllc}
\end{figure}

Reference sources are selected in a range of $\pm 0.2$~mag from the WISE flux of the TDF, with the exception of the brightest target (PTF-09ge), for which we use $\pm 0.8$~mag to ensure sufficient reference sources are available.  Reference sources that are not consistent with a constant flux are removed using a simple cut of $\chi^{2}_{r}>10$.  The final number of reference sources for each TDF is listed in Table~\ref{tab:photo}. 

Since each reference source is expected to have constant flux with time, the observed variability of the reference sources provides a measurement of the typical accuracy of the photometry for each target. We used Eq.~\ref{eq:relph} to find the relative photometry light curve of each reference source and computed the root-mean-square variability of this light curve, $\sigma(m_{{\rm rel},i})$. We list the mean of $\sigma(m_{{\rm rel},i})$ in the third column of Table~\ref{tab:photo}. For PTF-09ge, the brightest target in our sample, the photometric accuracy is 0.016~mag.

\begin{figure*}
\includegraphics[width=0.5 \textwidth]{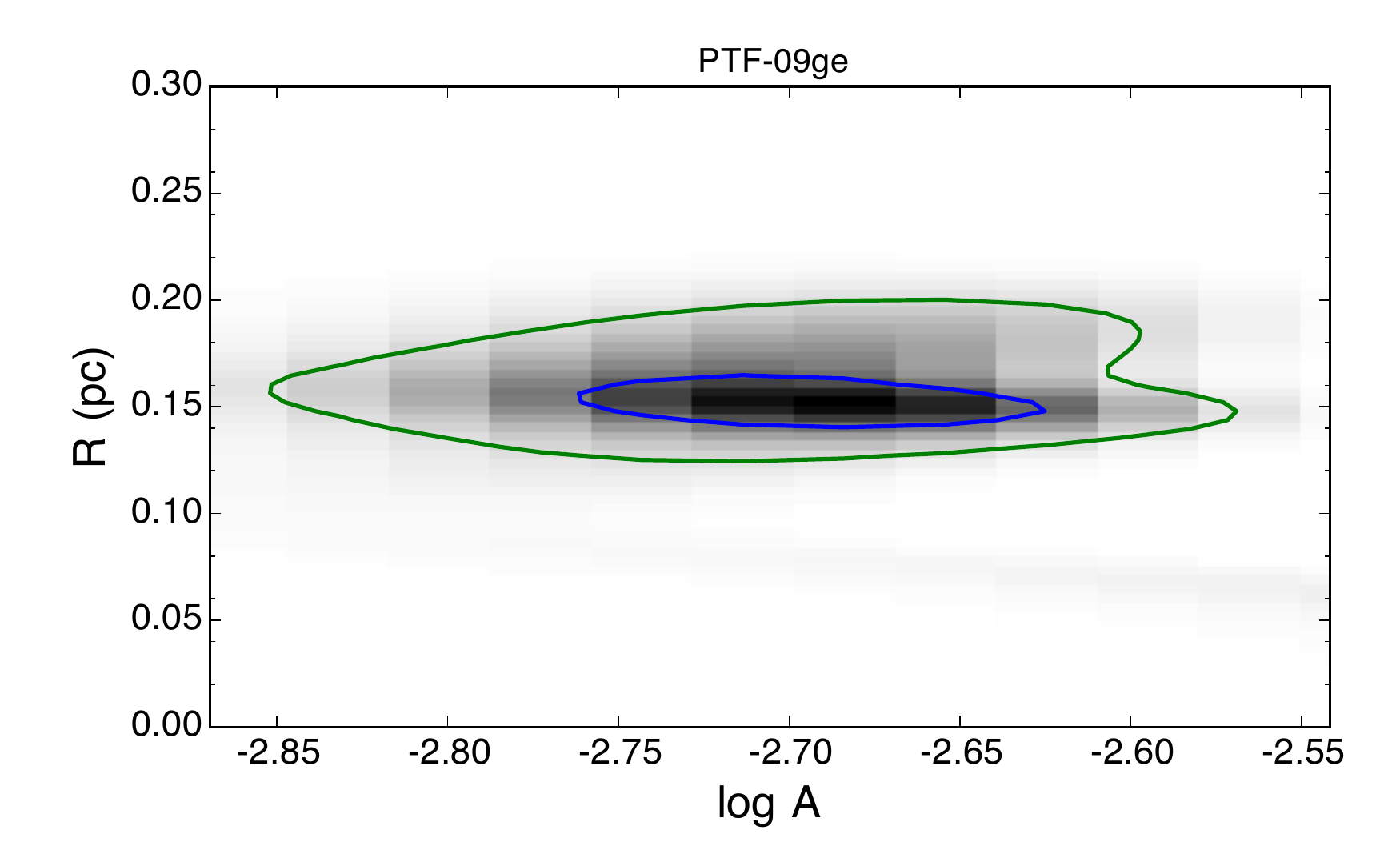}
\includegraphics[width=0.5 \textwidth]{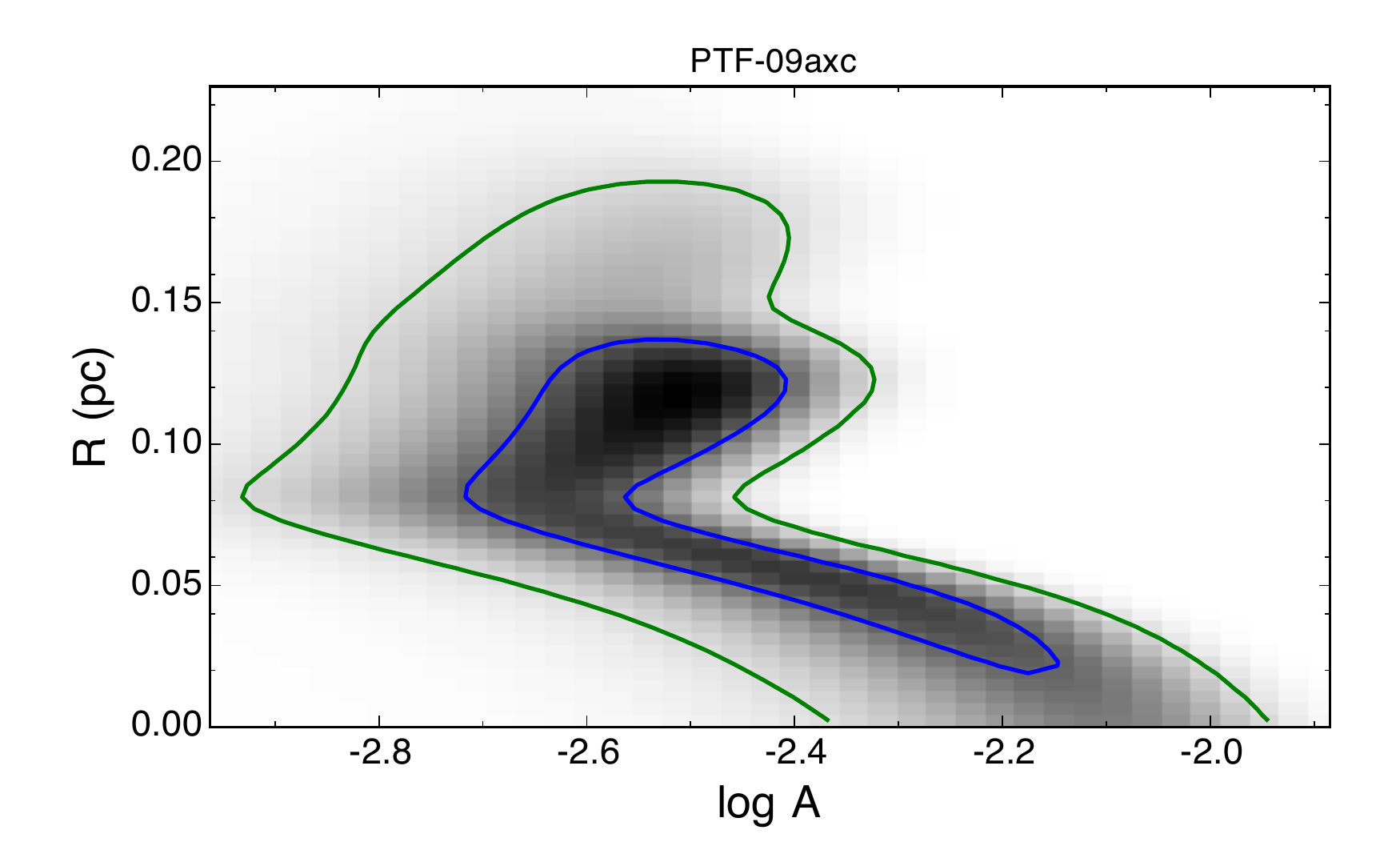}
\caption{Contours of constant likelihood obtained for our dust reprocessing model (Eq.~\ref{eq:rep}). Blue and green show the 68\% and 95\% confidence intervals, respectively. For PTF-09ge, a drop in the IR light curve that is steeper than the contemporaneous optical/UV light curve allows for an accurate determination of the shell radius, $R$. The light curve of PTF-09axc shows a more gradual decline that is broadly consistent with the shape of the optical light curve. This explains the non-vanishing probability at $R=0$, although this is not a physical solution within the reprocessing model.}\label{fig:2dlikel}
\end{figure*}

Three of the five TDFs in our sample show significant variability  ({i.e., the probability of a constant flux is less than 1\%, see~Table~2}) in the W1 band: PTF-09ge, PTF-09axc, and D23H-1. The signal-to-noise of the WISE observations of the other TDFs in our sample is not sufficient to detect variability at the level observed for these three flares. 

To find the difference flux, we use co-adds of 2014-15 to estimate the baseline host galaxy flux. A change in flux relative to this baseline is simply given by 
\begin{equation}\label{eq:Deltam}
\Delta m_{\rm TDF} = m_{\rm rel, TDF}(t) - \left<m_{\rm rel, TDF}(2014-15)\right> \quad.
\end{equation}
The uncertainty on $\Delta m_{\rm TDF}$ follows from adding the uncertainty of the baseline and the pre-2014 measurements in quadrature. The light curve of D23H-1 shows evidence for additional variability in 2015, hence $\Delta m_{\rm TDF}$ (Eq.~\ref{eq:Deltam}) is not well-defined for this source. In the following section we therefore focus only on PTF-09ge and PTF-axc, since these two flares show a large flux increase with respect to a relatively well-defined baseline of late-time observations. We note that PTF-09ge appears to show a modest decline in flux at late times, suggesting our estimate of the difference flux of this flare could be slightly ( $\sim$10\%) too low.

%
\section{Parameter inference}\label{sec:fit}
%
We now discuss how to estimate the parameters of our dust reprocessing model (Sec.~\ref{sec:theo}) from the observed light curve (Fig.~\ref{fig:alllc}). Our model assumes dust reprocessing in a thin shell, but this assumptions has little impact on our results (see Appendix). The results are summarized in Table~\ref{tab:results}. 

To compare our model to the observations we use an equivalent form to Eq.~\ref{eq:repdelta},
\begin{align}\label{eq:rep}
& L_{\nu, \, \rm model}(t, A, R, T) = \nonumber \\ 
&  \phantom{-} A~B'_{\nu}(T) ~2\pi \int_{0}^{\pi} {\rm d}\theta \, \sin \theta \, L_{\rm TDF}[t-\tau(R, \theta)] \quad.
\end{align}
The dust grains are not perfect black bodies and the IR spectrum of the dust is described by a modified Planck function:
\begin{equation}\label{eq:dustsed}
B_{\nu}'(T) = C_{q}\, B_{\nu}(T)\nu^{q} \quad, 
\end{equation}
with $q=1.8$ for grains of $a\sim 0.1$~$\mu$m and $1000<T/{\rm K}<2000$ \citep{Draine84} and $C_{q}$ a constant. For convenience, we normalize this spectrum such that the amplitude $A$ of the reprocessing signal is dimensionless (i.e., $\int {\rm d}\nu\, B_{\nu}'=C_{q}^{-1}$). The amplitude $A$ is determined from the observed IR luminosity; it measures the ratio between the observed IR luminosity and the amplitude of the reprocessing light curve, so we typically have $A \sim 10^{-3}$ (Fig.~\ref{fig:2dlikel}).

We use the well-sampled TDF PS1-10jh \citep{Gezari12} as a template for  $L_{\rm  TDF}(t)$. Since we are in the regime where $R/c$ is much larger than the typical timescale of the optical flare ($\Delta t_{\rm opt}$), the TDF energy determines the amplitude of the reprocessing signal, and the exact shape of the flare light curve is not important. The light curve of PS1-10jh provides a remarkably good match to PTF-09ge; only a small normalization shift (0.05~mag) and no temporal stretch have been applied. 
For the other flare, PTF-09axc, no late-time detections are available, and we will assume the post-flare light decays with the same rate as PTF-09ge. 

Although our IR light curves (Fig.~\ref{fig:alllc}) are sparsely sampled, the amplitude and radius can be accurately determined as they are nearly orthogonal when $\Delta t_{\rm opt}$ is shorter than $R/c$. In other words, our fiducial light curve model has very limited flexibility. 

We use a maximum likelihood method to find the best-fit value of $R$ and $A$. While we expect that the dust emits near the sublimation temperature ($T_{\rm d}=1850$~K), lower temperatures are possible if dust only exists at radii larger than the sublimation radius. To account for the uncertainty due to the unknown dust temperature, we use both the W1 and W2 observations and we allow dust temperatures in the range $T=[0, 1850]~K$. 
We maximize the logarithm of the likelihood given by:
\begin{align}\nonumber
 & \mathcal{L}(A,R, T) = \\   &  \phantom{-}  -\sum_i  \ln(\sigma_{i}) + [L_{\nu}(t_{i}) -L_{\nu, \, \rm model}(t_{i}, A,\,R,\,T)]^2 / 2\sigma_i^2  \quad,  
\end{align}
with $L_{\nu_{i}}$ the observed IR luminosity and the sum running over the observations in the W1 and W2 band.

Contours of constant likelihood are shown in Fig.~\ref{fig:2dlikel}. We see that the shell radius of \mbox{PTF-09ge} can be accurately  determined. For this flare, the probability at $R=0$ is vanishingly small, implying that the shape of the IR light curve is significantly different from the shape of the TDF light curve. For the other flare, PTF-09axc, an IR light curve that mirrors the TDF light curve can only be ruled-out at the 1$\sigma$ level.

To estimate the uncertainty on the relevant parameters, we sample the 3-dimensional likelihood using the straightforward Monte Carlo method of rejection sampling. We thus obtain distributions for sets of our three parameters ($A$, $R$, $T$) or for any scalar function that takes these parameters as input (i.e., $L_{\rm abs}$ and $f_{\rm dust}$). The 68\% confidence interval for a given parameter is then given by the 16 and 84 percentiles of the distribution of the parameter. 

Combinations of temperature and radius that yield $L_{\rm abs}<L_{\rm bb}$ (i.e., a total luminosity that is lower than the observed luminosity) are unphysical and excluded when computing the confidence intervals on $L_{\rm abs}$ and $f_{\rm dust}$. The temperature range that is excluded by the requirement $L_{\rm abs}<L_{\rm bb}$ is below the $2\sigma$ lower limit on the temperature based on the W2 observations. Hence this cut mainly serves to the remove unphysical solutions near $R=0$.

\begin{figure*}
\includegraphics[width=0.5 \textwidth]{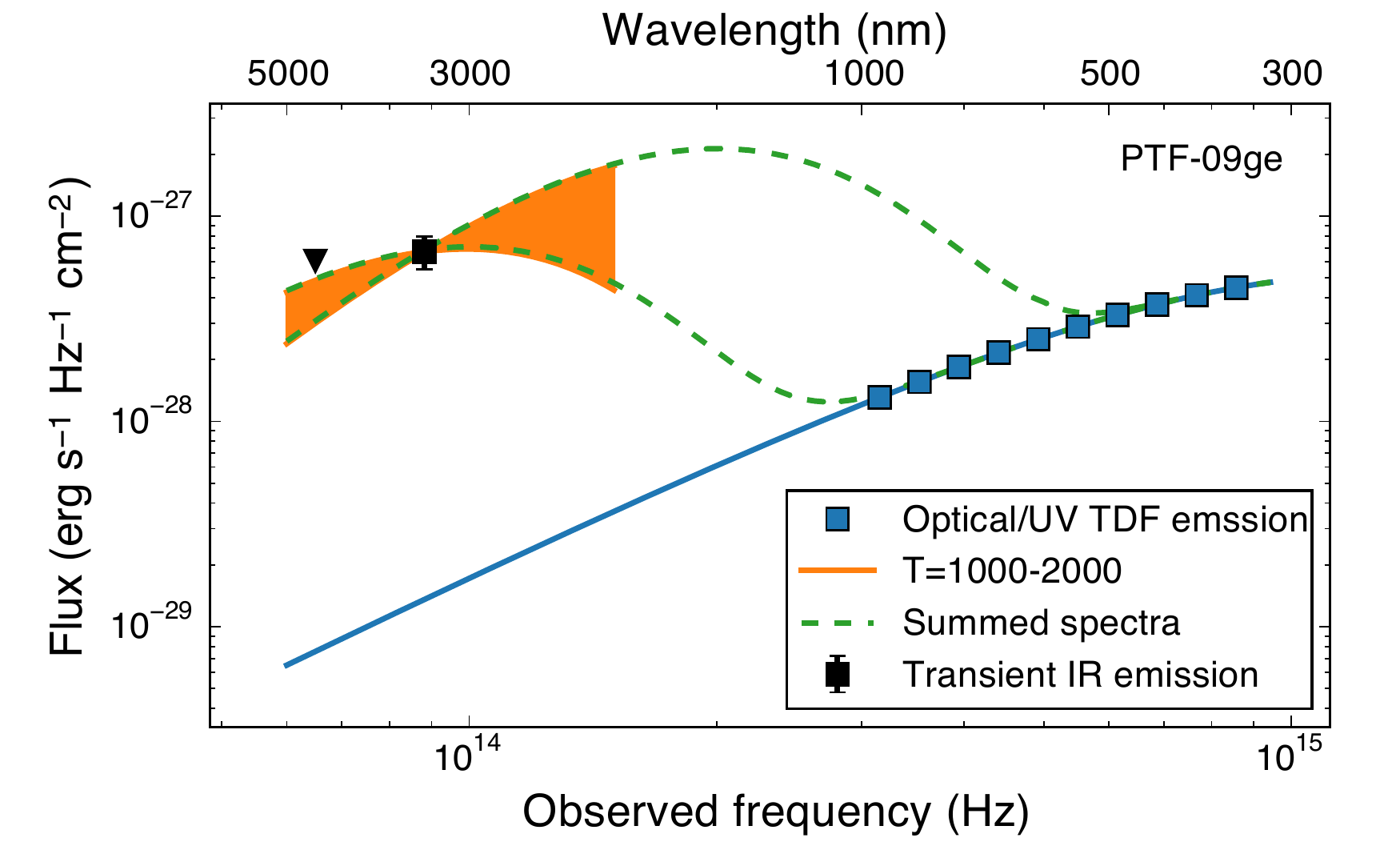}
\includegraphics[width=0.5 \textwidth]{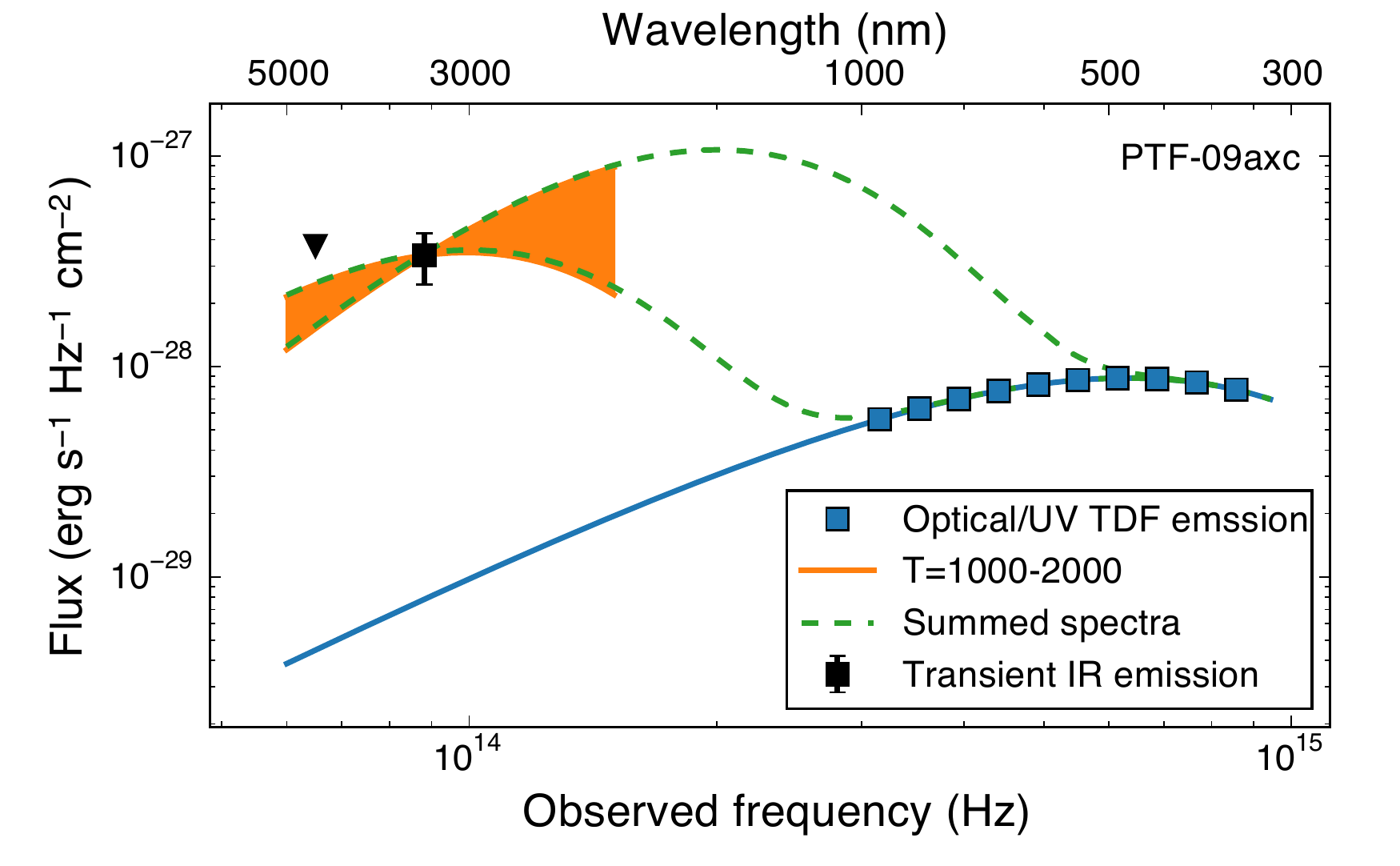}
\caption{Optical and IR SEDs at the time of the first WISE observation. The black symbols show the two WISE bands (W1 and W2, 2-$\sigma$ upper limits for the latter); the shaded region shows a modified black body (Eq.~\ref{eq:dustsed}), normalized to the W1 observations, for two different temperatures. The blue squares indicate the frequency range over which the TDF black body temperature has been measured. The amplitude of the TDF SED has been normalized to match the light curve at the time of the first WISE observation. We see that the expected W1 flux from the TDF SED  is orders of magnitude smaller than the observed WISE flux, implying that the IR emission originates from a different photosphere. The green dashed line shows the summed spectrum of the modified black body and the TDF SED. }\label{fig:sed}
\end{figure*}

%
%
\section{Discussion}\label{sec:results}
%
%

\subsection{Origin of the IR emission}\label{sec:origin}
We can rule out that the observed transient IR emission originates from the same photosphere as the optical/UV emission of the tidal flare. First, the shape of the IR light curve is significantly different from  the monotonic power-law decay that describes optical TDF emission. And second, the observed IR luminosity ($\nu L_{\nu}\sim 10^{42}\,{\rm erg}\,{\rm s}^{-1}$) is two orders of magnitude larger than the luminosity obtained from extrapolating the optical/UV black body spectrum, observed near the peak of the flare, to the time of the first WISE observation.  In Fig.~\ref{fig:sed} we show the optical black body spectra of the two TDFs in our final sample. Since the optical observations are on the Rayleigh-Jeans side of the black body spectrum and well-fit with a single temperature, we can extrapolate this SED to find the expected WISE flux from the TDF photosphere at the time of the WISE observations. For both flares, this extrapolated flux is at least an order magnitude lower than the observed flux. We can thus conclude that the IR emission does not originate from the optical/UV photosphere. To match the observed W1 flux with a black body of $T=2000$~K, the radius of this IR photosphere needs to be $\sim 10^{2}$ times larger than the radius of the optical/UV photosphere. 

At the time of the first WISE observations, emission from hot dust exceeds the TDF emission for wavelengths $\gtrsim700$~nm (see Fig.~\ref{fig:sed}) and therefore could be detectable with ground-based near-IR observations. 
For a generic dust reprocessing model, the IR emission is delayed, which could explain why for previous TDFs, a photosphere with $T\approx 2000$~K is not evident in optical spectra obtained near the peak of the flare.

Besides thermal emission from the TDF, a source of the observed transient IR flux could be synchrotron emission from a jet, such as observed for the relativistic TDF Swift~J1644+57 \citep{Burrows11}. However none of the TDFs in our sample have been detected in radio follow-up observations \citep{vanVelzen12b,Arcavi14}, ruling out emission from jets similar to Swift~J1644+57.  

With other potential sources of IR emission ruled out, dust reprocessing remains as the most plausible explanation for our observations. In fact, the light curve of \mbox{PTF-09ge} (Fig.~\ref{fig:alllc}) provides strong evidence for reprocessing by a spherical dust shell. As discussed in Section~\ref{sec:theo}, when a spherical shell is briefly illuminated by a central point source, the area that is seen to emit simultaneously to an observer at large distance is constant with time.

After two observations with a near-constant flux, the third point in the 3.4~$\mu$m light curve of \mbox{PTF-09ge}, 1.5~years after maximum light, shows a drop that is much steeper than the power-law decay of the optical TDF emission (Fig.~\ref{fig:alllc2}). At this point, the peak of the flare emission, which lasted only a few months, has been reprocessed and all of the resulting IR photons have crossed the shell. The time delay between the TDF peak and this drop provides a direct measurement of the radius where the reprocessing happens. For a spherical shell, we find $R=0.15_{-0.01}^{+0.03}$~pc (Sec.~\ref{sec:fit}).  

The observed infrared luminosity of the two flares in our final sample is an order of magnitude fainter than recent theoretical estimates of dust emission from TDFs by \citet{Lu16}. This difference is partially due to the dust covering factor that Lu et~al. adopted, which is an order of magnitude larger than the covering factor implied by our observations (see Sec.~\ref{sec:covering}).    

The light curve of the third flare with significant IR variability, D23H-1, cannot be explained by a single-shell reprocessing model because it shows a large change in flux at both 2 and 7 years after the optical peak. Future IR monitoring of this source is needed to measure the host galaxy baseline flux. We speculate that the IR light curve of D23H-1 can be explained by a concentration of dust at two different radii.  Alternatively, the late-time re-brightening of D23H-1 could be explained if the source of reprocessed emission is the accretion disk of an AGN (which could be tested with additional optical monitoring of this source). Finally, we note that the late-time light curve of PTF-09ge shows evidence for low-level emission from regions beyond the sublimation radius of graphite grains.

\begin{figure*}
\includegraphics[width=0.5 \textwidth]{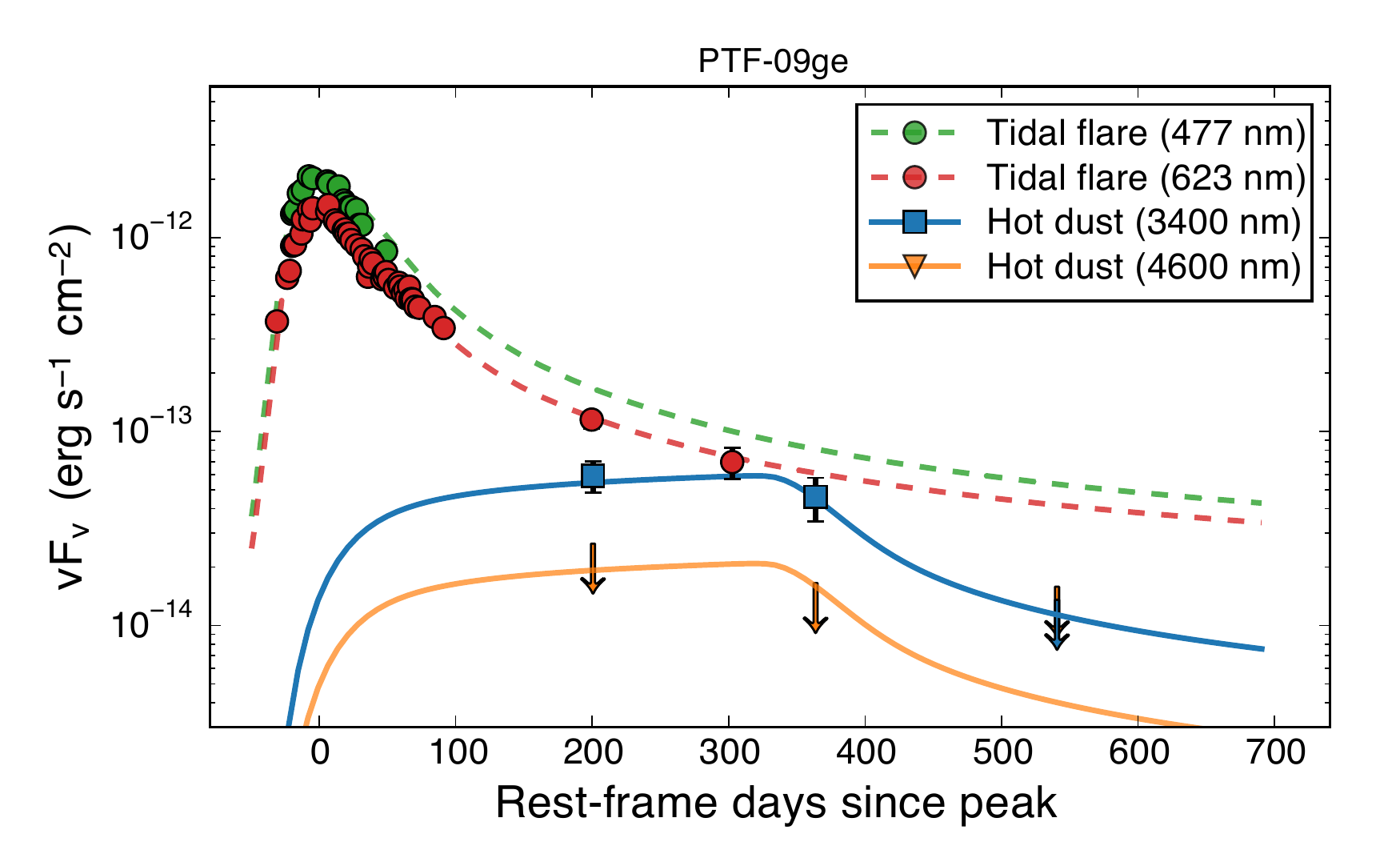}
\includegraphics[width=0.5 \textwidth]{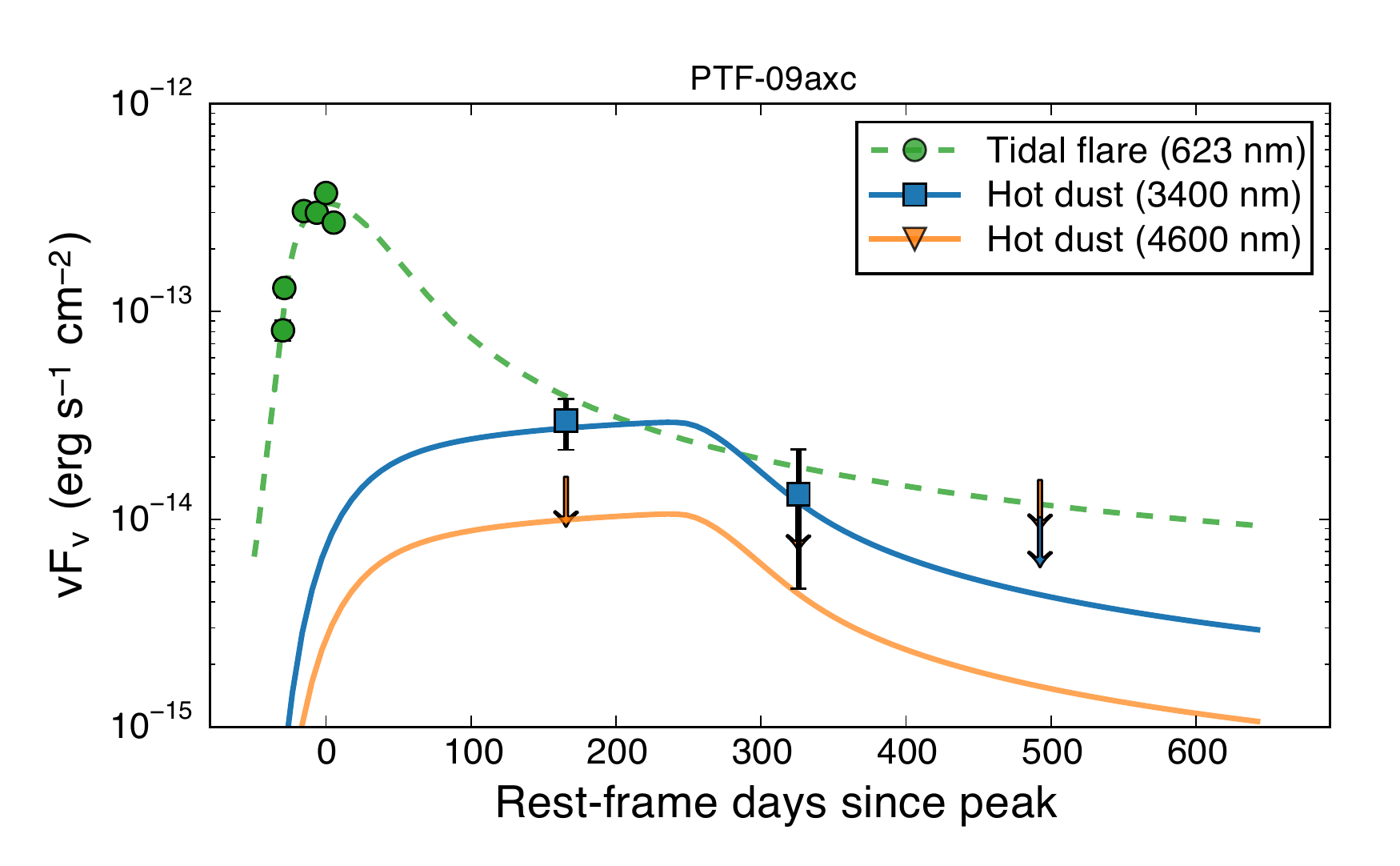}
\caption{Difference flux light curves of optical and IR emission TDFs. The circles show the optical observations of the flare; the dashed line that runs through these points is the light curve of the well-sampled tidal flare PS1-10jh. The baseline-subtracted IR data from WISE is shown by the square symbols; 1$\sigma$ upper limits are indicated by arrows. Our best-fit model for the IR emission, obtained by reprocessing the tidal flare light in a shell of hot dust ($T =1850$~K), is shown by the solid lines.}\label{fig:alllc2}
\end{figure*}

\subsection{TDF bolometric luminosity}\label{sec:TDFbol}
With the radius of the dust shell measured, we can estimate the luminosity of the flare emission that has carved out this shell ($L_{\rm abs}$, Eq.~\ref{eq:L_from_R}) in terms of the dust temperature. We anticipate the dust temperature is close to the sublimation temperature, $T_{\rm d} \approx 1850$~K, but our observations at 4.6~$\mu$m are not sensitive enough to provide a measurement of this temperature. This introduces an uncertainty on $L_{\rm abs}$ that is accounted for by marginalizing over temperature.
As discussed in detail in Section~\ref{sec:fit}, we use the observations at 3.4 and 4.6~$\mu$m to find the likelihood of the model light curve as a function of $T_{\rm d}$ and $R$. For the flare PTF-09ge, we find $\log[ L_{\rm abs} / {\rm erg}\, {\rm s}^{-1}] = 44.9^{+0.1}_{-0.6}$ (68\% confidence interval).

The TDF luminosity inferred from a dust reprocessing light curve closely approximates the bolometric flare luminosity because dust absorption is efficient from optical to soft X-ray frequencies, thus covering the full spectral range where thermal TDFs emit most of their energy --- including the extreme-UV which is nearly impossible to observe directly due to absorption by neutral hydrogen. The ratio between the luminosity inferred from the reprocessing light curve and the black body luminosity estimated from the optical observations is $0.8_{-0.5}^{+0.1}$ for PTF-09ge (logarithmic units). Either of two plausible explanations may account for this large bolometric correction\footnote{For AGN, the bolometric correction refers to the ratio between $\nu L_{\nu}$ at a given frequency and the total luminosity. Since  optical/UV observations of TDF are well-described by a black body spectrum, it is more instructive to define the TDF bolometric correction with respect to the black body luminosity.}.

First, the bolometric correction may be due to dust extinction in the host galaxy (i.e., due to dust along our line of sight to the black hole). This reddening decreases the observed black body temperature and, since the flare SED peaks at UV wavelengths, significantly reduces the inferred black body luminosity. The observed slope of the optical SED of PTF-09ge ($T=2.2\times 10^{4}~K$) is more shallow than the slope for the Rayleigh-Jeans limit, hence the intrinsic temperature of this flare could be higher than the observed temperature. Using the \citet{Calzetti00} extinction law to correct the observed optical spectrum of \mbox{PTF-09ge} for a dust column with $E_{B-V}=0.3$ increases the black body luminosity by a factor $\approx 8$. An analysis of sodium absorption in the spectra of type~1 AGN suggests that an extinction of $E_{B-V}= 0.3$ to a galactic nucleus is not uncommon \citep{Baron16}.

Alternatively, if the extinction to the center of the galaxy is small, the bolometric correction could be explained by adding an X-ray emitting component to the flare SED. The TDF ASASSN-14li \citep{Holoien15,Miller_14li,vanVelzen16} showed two black body spectra, with temperatures $T = 3\times10^{4}$~K  and $kT =  0.05$~keV. The latter dominates the total energy output and could thus explain the bolometric correction to the optical luminosity of PTF-09ge. 

If the accretion of stellar debris is radiatively efficient ($\eta\equiv L/\dot{m}c^{2}=0.1$), our estimate of the total radiated energy of \mbox{PTF-09ge} implies an upper bound on the accreted mass of $0.1 M_{\odot}$. Recent numerical simulations \citep{Shiokawa15} show a similar mass accretion after the disruption of a solar type star, hence our observations are consistent with high radiative efficiency in super-Eddington accretion disks \citep{Jiang14}. This inference also points to a full disruption of a solar-type star, while a partial disruption would be inferred if no bolometric correction is applied to the light curve. 

\begin{deluxetable*}{l cccccc}
\tablewidth{0.8 \textwidth}
\tablecolumns{6}
\tablecaption{{Parameters derived from IR observations}\label{tab:results}}
\tablehead{Name & $\log L_{\rm IR}$  & $R$ & $\log L_{\rm abs}$ &  $\log E_{\rm abs}$  &$\log f_{\rm dust}$ \\ 
 & $({\rm erg}\,{\rm s}^{-1})$ & $({\rm pc})$ & $({\rm erg}\,{\rm s}^{-1})$ &  $({\rm erg})$   & }\\
 \startdata

PTF-09ge   & $41.8 \pm 0.1$ & $0.15_{-0.01}^{+0.03}$ & $44.9_{-0.6}^{+0.1}$ & $52.0_{-0.6}^{+0.1}$ & $-2.0_{-0.2}^{+0.2}$\\
PTF-09axc  & $42.0 \pm 0.1$ & $0.12_{-0.04}^{+0.07}$ & $43.9_{-0.2}^{+0.8}$ & $51.1_{-0.2}^{+0.8}$ & $-1.6_{-0.5}^{+0.4}$\\

\enddata
\tablecomments{For the two tidal flares with transient 3.4~$\mu$m flux, we list the observed peak IR luminosity in the second column. The third to last columns show the results of applying our dust reprocessing model to the IR light curve. From the break in the light curve we find the radius of the reprocessing shell ($R$). For a given dust temperature, we can estimate the peak luminosity of the tidal flare over the frequencies where dust absorbs the flare photons ($L_{\rm abs}$). Integrating $L_{\rm abs}$ over the flare light curve yields the total radiated energy ($E_{\rm abs}$).  Finally, the covering factor ($f_{\rm dust}$) is given by the total energy radiated by the dust divided by $E_{\rm abs}$. We list the parameters that correspond to the maximum likelihood and the uncertainties reflect the 68\% confidence interval. This interval includes a marginalization over the dust temperature, which provides the largest contributions to the uncertainty.}
\end{deluxetable*}

\subsection{Dust covering factor}\label{sec:covering}
Our observations are the first to probe dust within 0.1~parsec of the center of non-active galaxies. We can use the total energy radiated in the IR ($E_{\rm dust}$) and the energy able to heat graphite dust ($E_{\rm abs}$) to find the covering factor of this dust, $f_{\rm dust} = E_{\rm dust} / E_{\rm abs}$. For both TDFs in our final sample we find $f_{\rm dust}\sim 1$\% (Table~\ref{tab:results}).

The fractional uncertainty on $f_{\rm dust}$ is smaller than for $L_{\rm abs}$ since the former has a weaker temperature-dependence. The ratio between the total IR luminosity and the portion we observe at 3.4$\mu$m (cf. Eq. \ref{eq:Ldust}), increases with temperature, $\propto T^3$ for temperatures $\simeq 1800$~K, but somewhat less steeply at lower temperatures. Since $E_{\rm abs}\propto T^{5.8}$ (Eq.~\ref{eq:L_from_R}), the fractional uncertainty on the covering factor is a factor $\approx2$ smaller than the fractional uncertainty on $L_{\rm abs}$. 

Galaxy-to-galaxy fluctuations in the dust size distribution are expected to have a small influence on the inferred covering factor. Dust lanes in E/S0 galaxies have an extinction curve that is similar to the Milky Way \citep{Finkelman12}, which implies a similar peak of the grain size distribution \citep{Goudfrooij94}. For a sample of 26 early-type galaxies, the mean grain size difference with respect to the Milky Way is 8\% \citep{Patil07}. Since $E_{\rm abs} \propto a^{2}$ (Eq.~\ref{eq:L_from_R}), this fluctuation of the grain size translates to an uncertainty of 0.1~dex on $f_{\rm dust}$. 

While the dust distribution at the centers of galaxies is not constrained by observations, most mechanisms that can alter the distribution (e.g., sputtering) will act to reduce the number of small grains relative to large grains and therefore not affect our estimate of the covering factor, unless there are also agglomerative mechanisms particular to galactic nuclei. 

The information contained in the IR light curve (Fig.~\ref{fig:alllc2}) is not sufficient to constrain to what degree the dust geometry departs from spherical symmetry (see Appendix). However this does not affect our ability to measure the covering factor because this parameter is a measure of the absorbed energy, which is independent of the dust geometry (this is demonstrated in Fig.~\ref{fig:modelcompare}, right panel). 


The covering factor inferred from our observations is almost two orders of magnitude smaller than typical dusty tori in Seyfert galaxies \citep{Barvainis87}. This is not surprising since a high accretion rate is likely required to build and sustain a torus that covers a large solid angle \citep[e.g.,][]{Pier92}, while the host galaxies of the TDFs show no signs of high accretion rates prior to the stellar disruption. The Galactic Center likely provides a better comparison. Interestingly, the nuclear dust in the TDF host galaxies is different from the circumnuclear ring of molecular gas at the Galactic Center; this ring has a covering factor of about 20\% and a sharp inner edge at $\approx1.5$~pc from Sgr~A*  \citep[][]{Genzel10,Ferriere12}. Inside this edge, a region known as the ``ionized cavity'', free floating dust particles will be sublimated by the UV radiation of the nuclear star cluster. Yet dust can exist within dense molecular cores that are observed inside the cavity \citep{Christopher05}. Similar clumps may be the source of nuclear dust in the TDF host galaxies. Alternatively, if the TDF host galaxies have a nuclear star cluster with an old stellar population or lack a nuclear star cluster, dust on 0.1~pc scales may simply originate from streams of cold gas that fall toward the central black hole without being ionized.

Finally, we point to a potential section effect which could explain the relatively small covering factor of the galaxies in our sample. Since the optical/UV SED of the TDF candidates found so far is relatively blue ($T \sim 3 \times10^{4}$~K), extinction along the line of sight will quickly reduce their detectability. While color is not used as a selection criterion to find TDF candidates in optical transient surveys \citep[e.g.,][]{vanVelzen10,Arcavi14}, these surveys might still have a moderate bias to finding flares in galaxies with a small amount of dust. If the amount of dust along our line of sight to the galaxy center is strongly correlated with the amount of circumnuclear dust on sub-parsec scales, a TDF sample could yield a biased view of the nuclear dust covering factor. For a large sample of TDFs, this potential bias could be quantified by comparing the dust covering factor, as measured from the reprocessing light curve, to other dust extinction estimates (e.g., as obtained from the Balmer decrement or stellar population synthesis).

\section{Conclusions and Outlook}\label{sec:con}
Our main conclusions are:
\begin{itemize}
\item We have discovered variable 3.4~$\mu$m emission for three of the five tidal disruption flares in our sample (Fig.~\ref{fig:alllc2}), the host-subtracted IR luminosity is $\nu L_{\nu} \sim 10^{42}\,{\rm erg}{\rm s}^{-1}$. 
\item For two of the five TDFs in our sample, the observed IR light curves can be explained by reprocessing of UV/X-ray emission from the flare in a thin shell at $\sim 0.1$~pc from the black hole (Fig.~\ref{fig:alllc2}).
\item The radius of the IR-emitting region yields an estimate of the bolometric luminosity of the TDF. We find a typical luminosity of $10^{45}$~{erg}\,{s}$^{-1}$ (Table~\ref{tab:results}), which is a factor $\sim 10$ larger than the observed black body luminosity. 
\item The covering factor of the nuclear dust that is responsible for reprocessing the TDF light is $\sim 1$\% (Table~\ref{tab:results}).
\end{itemize}
For a dust temperature of $T \sim 2000$~K, ground-based IR observations of newly discovered TDFs could be used to estimate the dust temperature and obtain reprocessing light curves at higher cadence than presented in this work. These observations would reduce the uncertainty on the TDF bolometric luminosity and will also provide information on the geometry of the nuclear dust. High-resolution spectroscopic observations (e.g., with JWST) can be used to study in detail the response of dust particles to a sudden burst of UV radiation. 

\acknowledgments
We thank B. Ochsendorf and K. Tchernyshyov for useful discussions. We thank the referee for the useful comments. This publication makes use of data products from the Wide-field Infrared Survey Explorer, which is a joint project of the University of California, Los Angeles, and the Jet Propulsion Laboratory/California Institute of Technology, funded by the National Aeronautics and Space Administration. S.v.V. is supported by NASA through a Hubble Fellowship (HST-HF2-51350).

\bibliography{/Users/sjoertvanvelzen/Documents/articles/general_desk.bib}

\bibliographystyle{apj}

\clearpage
\appendix

%
\section{Alternative dust geometries}
%

Our fiducial model (Eq.~\ref{eq:rep}) assumes emission from a single shell. In this section we explore to what extent deviations from this assumption are constrained by the data and how they affect the inferred parameters.

We first consider a superposition of shells at different radii. The reprocessing light curve for this geometry follows by integrating Eq.~\ref{eq:rep} over $R$,
\begin{equation}\label{eq:sphere}
L_{\nu, \rm sphere}(t, R, T) = A ~2\pi \int_{R_{0}}{\rm d}R ~ n(R) B'_{\nu}[T(R)] \int_{0}^{\pi}{\rm d}\theta ~ \sin \theta \, L_{\rm TDF}[t-\tau(R, \theta)] \quad.
\end{equation}
Here $n(R)$ is the number density of the dust as a function of radius and $T(R)= T_{0} (R/R_{0})^{-2/5.8}$, with $T_{0}$ the temperature of the inner shell (cf. Eq.~\ref{eq:L_from_R}). Eq.~\ref{eq:sphere} does not account for self-shielding of the dust, which is justified as long as the covering factor is small. The width of the response function of each shell  (Eq.~\ref{eq:repdelta}) is proportional to the shell radius, hence the amplitude of the reprocessing light curve decreases with $n(R)/R$, multiplied by a small spectral correction of order $T(R)/T_{0}= (R/R_{0})^{-0.3}$. As a result, the IR light curve for a constant dust density has a much more shallow flux decay at $t=2R_{0}/c$ compared to a single shell. The light curve of \mbox{PTF-09ge} appears inconsistent with a constant density (Fig.~\ref{fig:modelcompare}, left panel). Using only the W1 observations, the log likelihood difference compared to the fiducial model is $-3.4$. 

Clearly, our observations have too few degrees of freedom to measure a slope of a dust density profile, $n(R)\propto R^{\alpha}$, but we can conclude that steep profiles ($\alpha\lesssim -1$) provide a better description of the data for PTF-09ge. For example, a Bondi density profile ($\alpha=-3/2$) with an inner radius at $R_{0}=0.1$~pc yields a reasonable fit (log likelihood difference with respect to the fiducial model is -1.4).

Finally, we also consider a disk geometry for the dust distribution. For a thin disk, the reprocessing light curve is given by 
\begin{equation}\label{eq:disk}
L_{\nu, \rm disk}(t, R, T) = A\, B'_{\nu}(T) \int_{0}^{2\pi}{\rm d}\phi ~  L_{\rm TDF}[t-\tau(R, \theta({\phi}))] \quad.
\end{equation}
Here $\theta({\phi})$ parametrizes the coordinates of disk, which run between $\theta =[i-\pi/2, \pi/2-i]$, with $i$ the inclination of the disk (e.g., for a face-on disk, $i=\pi/2$, all reprocessed emission has the same time delay of $R/c$). Our observations cannot disentangle the degeneracy between radius and inclination of a disk geometry. Yet we can use Eq.~\ref{eq:disk} to fit for the radius of the disk for a range of inclinations and then use Eq.~\ref{eq:L_from_R} to estimate the absorbed luminosity and the dust covering factor for these radii. We find that the range of inclinations that are consistent with the data of \mbox{PTF-09ge} yield a similar value of the covering factor compared to our fiducial dust model (Fig.~\ref{fig:modelcompare}, right panel). This should not be a surprise since the covering factor measures an energy ratio, i.e., it has been integrated over the geometry of the emitting region.

\begin{figure*}
\includegraphics[width=0.50 \textwidth]{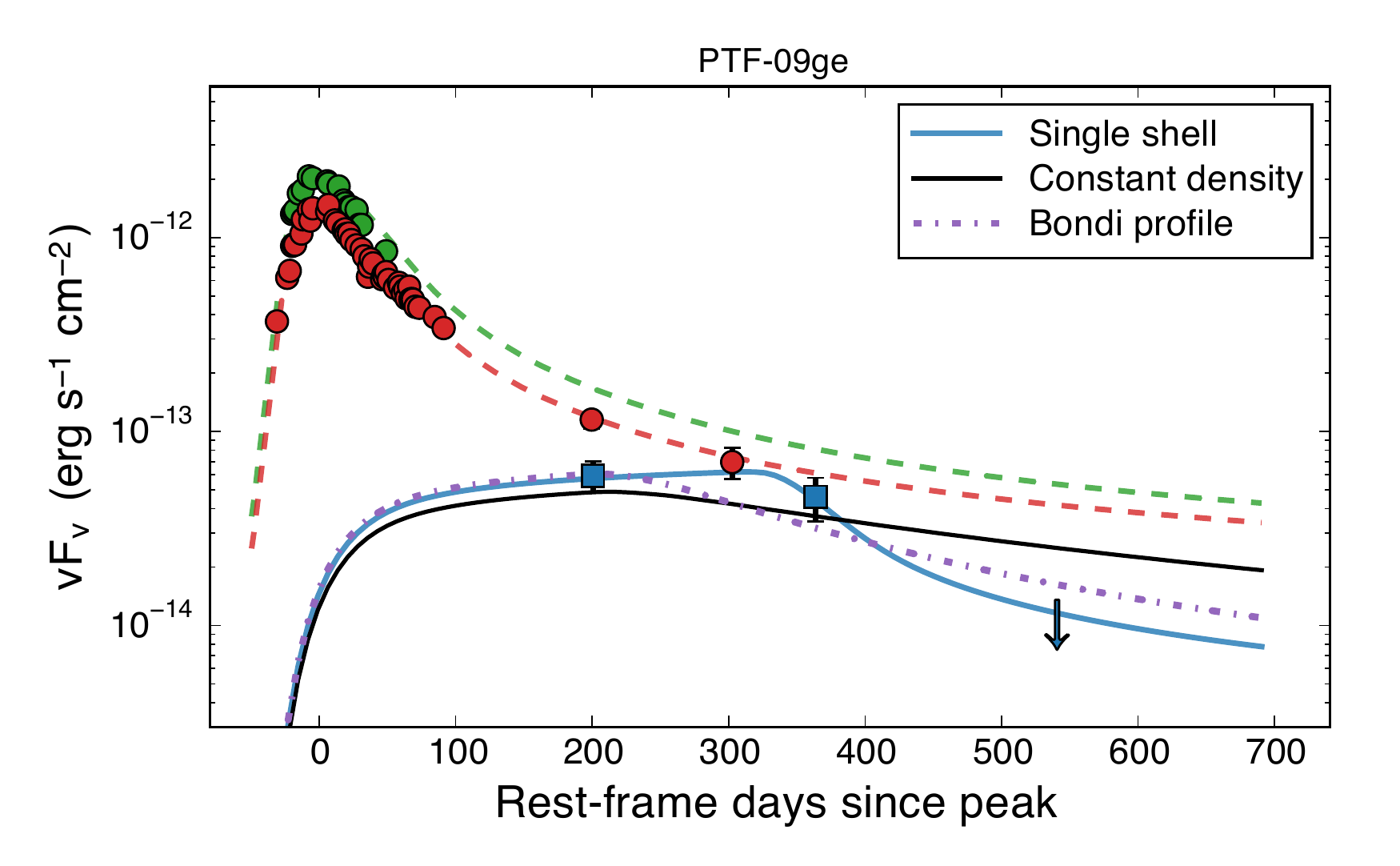} 
\includegraphics[width=0.50 \textwidth]{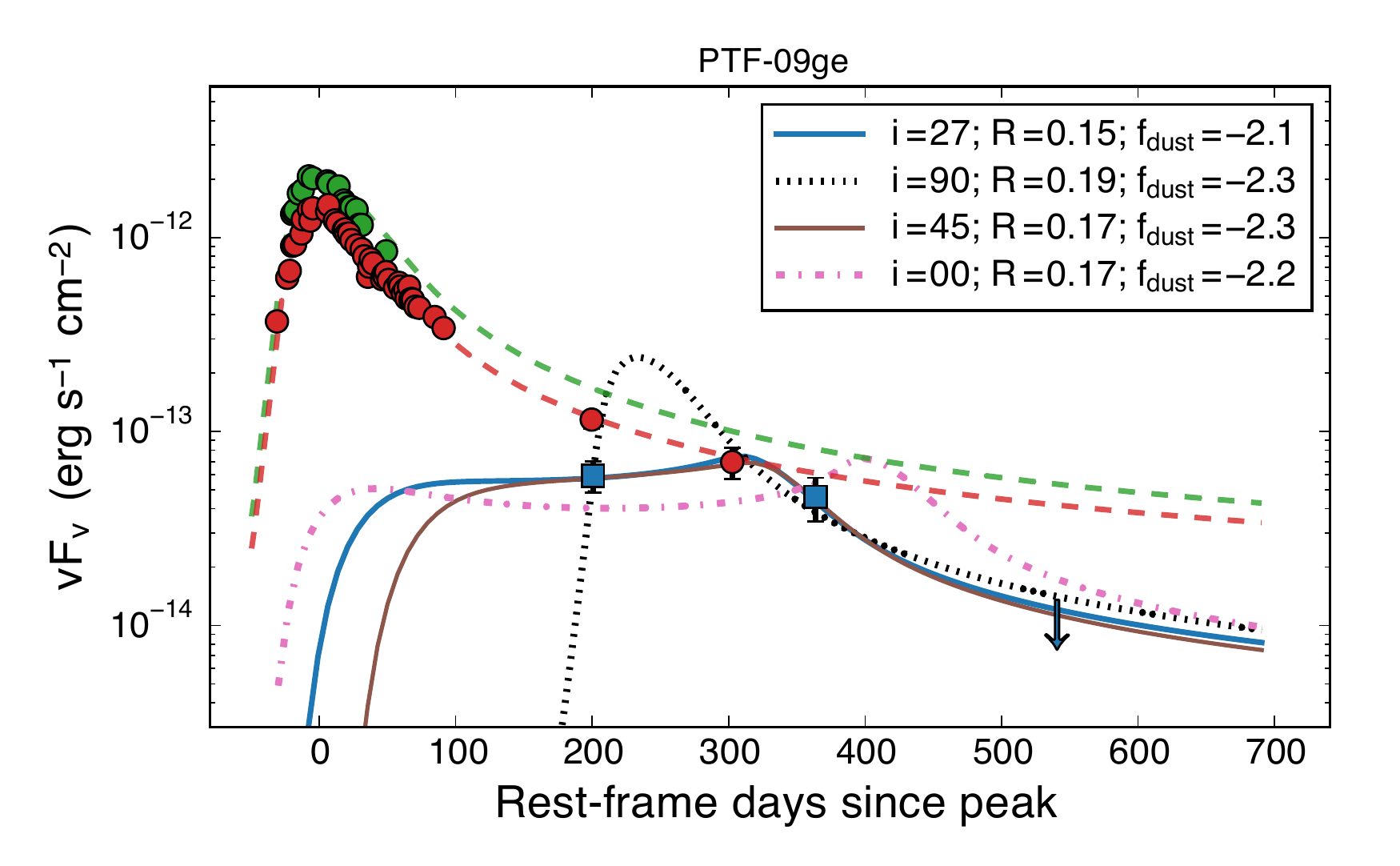}
\caption{Reprocessing light curves of \mbox{PTF-09ge} for alternative dust geometries (legend identical to Figure~\ref{fig:alllc2}, with the exception of the different reprocessing light curves). {\it Left:} our fiducial single-shell model (blue solid line) and the IR light curves for two different radial density profiles of the dust: a constant density (black solid line) and a Bondi density profile (dashed-dotted line). The constant density profile does not provide a good description of the data. {\it Right:} the reprocessing light curve for four different disk geometries of the dust. The inclination (in degree), radius (in parsec) and $\log$ of the covering factor are listed in the legend. For the first model (solid blue line), we kept the dust radius fixed at the radius measured for our fiducial spherical shell model and find the best-fit inclination ($i=27$~deg). While for the other three models, we fixed the inclination and measured the best-fit radius of the dust shell. Except for an edge-on disk ($i=0$), all disk models appear consistent with the data and all yield a dust covering factor that is very similar to the fiducial value ($\log f_{\rm dust}=-2.0$) obtained for a spherical dust shell.}\label{fig:modelcompare}
\end{figure*}

\end{document}